\documentclass[journal,twoside]{IEEEtran}
\usepackage{graphicx,hyperref,amsmath}
\newcommand{\etal}{{\it et~al.\/} }

\begin{document}
%
% paper title
\title{Study of the Dynamics of Large Reflector Antennas with Accelerometers}

%
%
% author names and IEEE memberships
% note positions of commas and nonbreaking spaces ( ~ ) LaTeX will not break
% a structure at a ~ so this keeps an author's name from being broken across
% two lines.
% use \thanks{} to gain access to the first footnote area
% a separate \thanks must be used for each paragraph as LaTeX2e's \thanks
% was not built to handle multiple paragraphs
\author{Ralph~C.~Snel, %~\IEEEmembership{Member,~IEEE,}
        Jeffrey~G.~Mangum, %~\IEEEmembership{Fellow,~OSA,}
        and~Jacob~W.~M.~Baars%~\IEEEmembership{Life~Fellow,~IEEE
\thanks{Manuscript received July 10, 2006; revised November 30,
  2006.}%
\thanks{The performance results presented in this publication were
  part of a comprehensive technical evaluation process used to
  evaluate the ALMA prototype antennas which concluded in April 2005.}}

%\thanks{M. Shell is with the Georgia Institute of Technology.}}
% note the % following the last \IEEEmembership and also the first \thanks - 
% these prevent an unwanted space from occurring between the last author name
% and the end of the author line. i.e., if you had this:
% 
% \author{....lastname \thanks{...} \thanks{...} }
%                     ^------------^------------^----Do not want these spaces!
%
% a space would be appended to the last name and could cause every name on that
% line to be shifted left slightly. This is one of those "LaTeX things". For
% instance, "A\textbf{} \textbf{}B" will typeset as "A B" not "AB". If you want
% "AB" then you have to do: "A\textbf{}\textbf{}B"
% \thanks is no different in this regard, so shield the last } of each \thanks
% that ends a line with a % and do not let a space in before the next \thanks.
% Spaces after \IEEEmembership other than the last one are OK (and needed) as
% you are supposed to have spaces between the names. For what it is worth,
% this is a minor point as most people would not even notice if the said evil
% space somehow managed to creep in.
%
% The paper headers
\markboth{IEEE Transactions on Antennas and Propagation Magazine,~Vol.~49,
  No.~4,~August~2007,~pp.~84-101}{Snel \MakeLowercase{\textit{et al.}}: Study of the Dynamics of Large Reflector Antennas with Accelerometers}
% The only time the second header will appear is for the odd numbered pages
% after the title page when using the twoside option.
% 
% *** Note that you probably will NOT want to include the author's name in ***
% *** the headers of peer review papers.                                   ***

% If you want to put a publisher's ID mark on the page
% (can leave text blank if you just want to see how the
% text height on the first page will be reduced by IEEE)
\pubid{1045--9243/00\$00.00~\copyright~2007 IEEE}

% use only for invited papers
%\specialpapernotice{(Invited Paper)}

% make the title area
\maketitle

\begin{abstract}
%\section{Summary}
%================
%\label{summary}

The Atacama Large Millimeter Array (ALMA) will consist of up to 64
state-of-the-art sub-mm telescopes, subject to stringent performance
specifications which will push the boundaries of the technology, and
makes testing of antenna performance a likewise challenging task. Two
antenna prototypes were evaluated at the ALMA Test Facility at the Very
Large Array site in New Mexico, USA. The dynamic behaviour of the
antennas under operational conditions was investigated with the help
of an accelerometer system capable of measuring rigid body motion of
the elevation structure of the antenna, as well as a few low-order
deformation modes, resulting in dynamic performance numbers for
pointing stability, reflector surface stability, path length stability,
and structure flexure. Special emphasis was given to wind effects,
one of the major factors affecting performance on
timescales of seconds to tens of minutes.

Though the accelerometers could not directly measure antenna performance
on timescales longer than 10 seconds, it was possible to use them to
derive antenna properties which allowed extrapolation of the wind-affected
performance to timescales of 15 minutes and longer. This paper describes
the accelerometer system, its capabilities and limitations, and presents
the dynamic performance results of the two prototype antennas investigated.

In addition to verification of the performance requirements, we
investigated the vibration environment on the antennas, relevant for
vibration-sensitive equipment for use on the ALMA antennas, the lowest
eigenfrequencies for the antennas, and the sensitivity to vibration
generated by similar antennas operating nearby.

This work shows that seismic accelerometers can successfully be used
for characterisation of antenna dynamics, in particular when complemented
with simultaneous wind measurements and measurements for static
performance. The versatility of the accelerometers
also makes them a valuable tool for troubleshooting of unforeseen
antenna features.

\end{abstract}


\section{Introduction}
%================
\label{intro}

The enormous growth of radio astronomy in the millimeter and
submillimeter wavelength regime (frequencies from 100 - 1000 GHz) over
the last 25 years has been made possible both by the emergence of
sensitive detectors like SIS-diodes and HEB-devices and the
construction of ever larger and more accurate radio
telescopes. Reflector antennas for this wavelength region have been
built from relatively small (up to 15 m diameter), extremely accurate
(reflector accuracy \mbox{15 - 25 $\mu$m}) submillimeter telescopes to larger
(20 - 45 m) millimeter antennas with surface accuracy from \mbox{75 - 150
$\mu$m}. 

Current projects in this area are the 50 m diameter Large Millimeter
Telescope (LMT) under construction in Mexico \cite{Kaercher2000},
which is aiming to reach \mbox{75 $\mu$m} rms surface and 1 arcsec
pointing accuracy and the Atacama Large Millimeter Array (ALMA). ALMA
is a global collaboration of North America (USA and Canada) and Europe
(European Southern Observatory) with contributions from Japan to build
a powerful millimeter wavelength aperture synthesis array at a 5000 m
high plateau in Northern Chile. The instrument will consist of up to
64 high accuracy Cassegrain reflector antennas of 12 m diameter with a
surface rms accuracy of \mbox{20 $\mu$m} and a pointing and tracking
accuracy and stability of 0.6 arcseconds, all under the severe
operational conditions of the high site. This telescope will operate
over the entire frequency range from 30 to 950 GHz. 

These specifications are among the most severe ever realised in radio
telescopes and force the designer and manufacturer to push the
boundaries of the technology. At the same time, it is becoming
increasingly difficult for the contractor and the customer to
quantitatively and reliably evaluate the performance characteristics
of these instruments. At the longer wavelengths radio astronomers have
developed and used methods of antenna evaluation which are based on
the use of strong cosmic radio sources and astronomical observing
techniques \cite{Baars1973}. These measurements are only of limited use
at the short millimeter wavelengths at frequencies above 100 GHz. The
number of suitable cosmic test sources is severely limited because of
their intrinsic weakness and the sensitivity limitations of the relatively small
antennas. 

% The following will force second column to "pull-up" and allow pubid
% to be printed...
\pubidadjcol 

\begin{figure}
\resizebox{\hsize}{!}{
%  \centering
%  \includegraphics[scale=0.48,angle=0]{/Users/jmangum/Pictures/MyPictures/ATF/VertexRSI/Vertex.jpg}}
  \includegraphics[scale=0.48,angle=0]{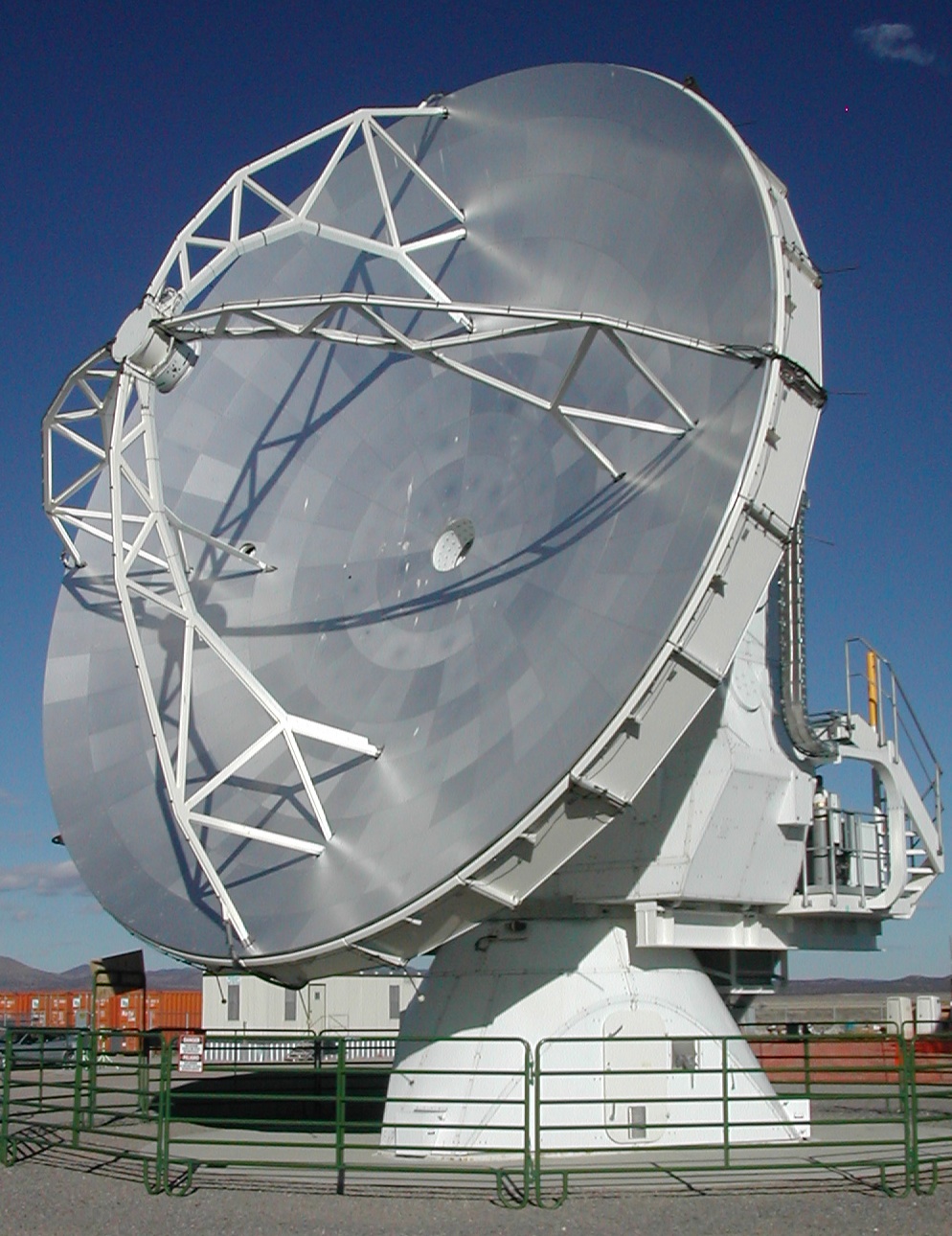}}
  \caption{The VertexRSI ALMA prototype antenna.}
  \label{fig:vertexrsi}
\end{figure}

\begin{figure}
\resizebox{\hsize}{!}{
  \includegraphics[scale=0.65,angle=0]{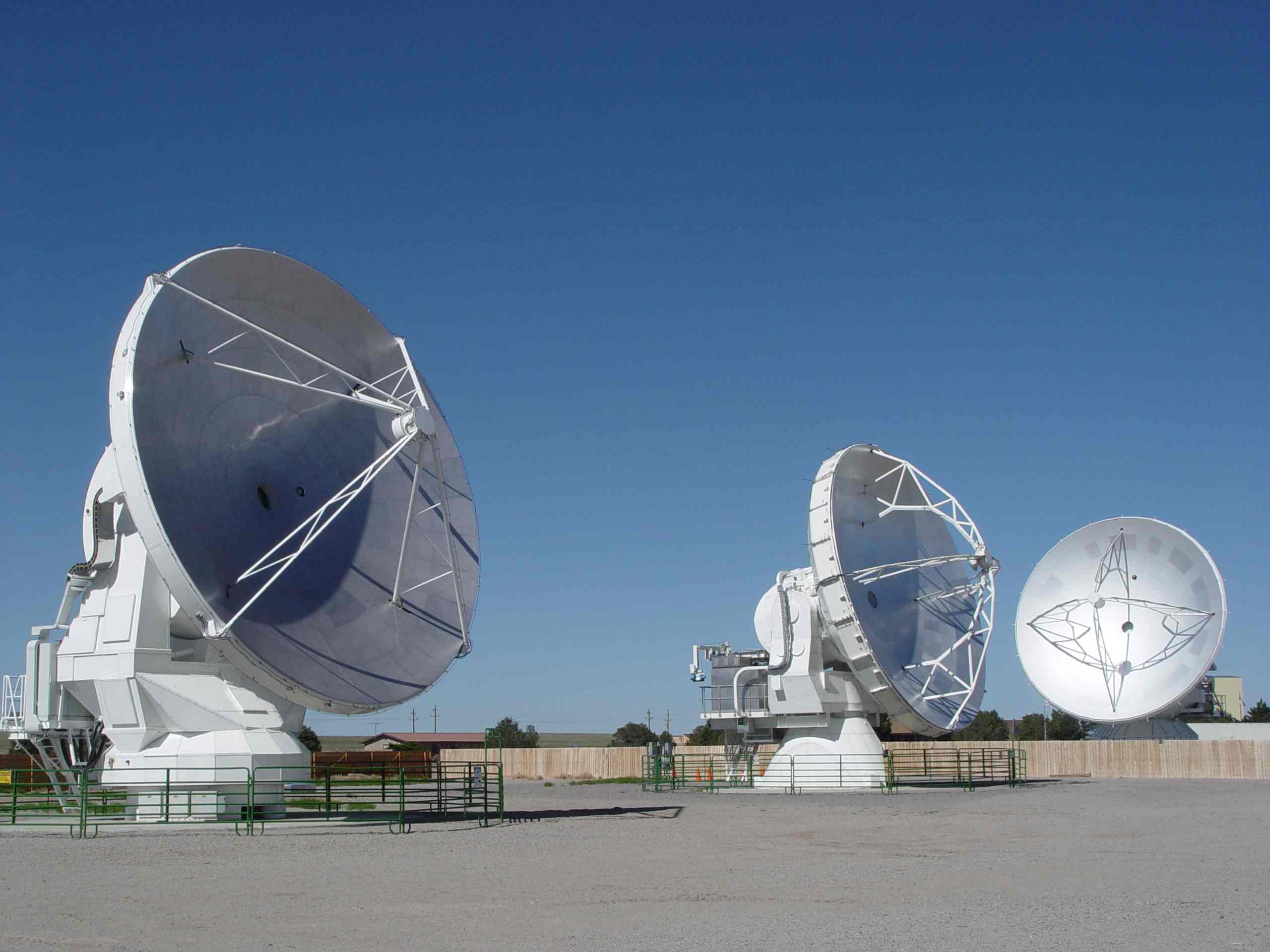}}
  \caption{The ALMA Test Facility, with the AEC antenna (left), VertexRSI antenna (middle) and Mitsubishi antenna (right).}
  \label{fig:atf}
\end{figure}

For the ALMA Project the partners decided early to obtain two
prototype antennas from different design and fabrication groups to
increase the chance of achieving the desired performance. The
prototype antennas tested here, one designed and constructed by
VertexRSI, the other by a consortium of Alcatel and European
Industrial Engineering, hereafter referred to as AEC, are similar in
overall design and built to meet identical requirements, though
significant differences exist in the approaches taken to meet
these requirements. The antennas, together with a third one from
Japan, are located 
at the ALMA Test Facility (ATF), on the site of the Very Large Array
(VLA) in New Mexico, USA (Figures~\ref{fig:vertexrsi} and
\ref{fig:atf}). For more information on the antennas and the
evaluation program, see \cite{Mangum2006}.

An international team of radio astronomers was formed to subject the
antennas to an extensive evaluation program. In preparing their tasks
this group determined that additional measurements and test methods
and instruments beyond the usual astronomical testing would be needed
to check the very stringent specification of the antennas. A
particularly important, but difficult to measure quantity is the
accuracy with which the antenna can be pointed at arbitrary positions
on the sky and the stability with which such a pointing can be
maintained under the influence of variations in temperature and wind
forces. Given our need to check these parameters independently of the
availability of celestial radio sources, we looked into the
possibility of using accelerometers on the antenna structure to
establish its dynamical behaviour. 

The use of seismic accelerometers for performance characterization has
been successfully demonstrated on optical telescopes \cite{Smith2004}
and mm-antennas \cite{Ukita2002}, \cite{Ukita2004}. 
Using a set of 10 seismic accelerometers, installed on the antenna
back-up structure 
(BUS), subreflector support structure (apex), and receiver cabin, we
have measured accelerations
allowing determination of rigid body motion of the elevation
structure, and a few low-order distortions of the BUS.  

The nature of the accelerometers used in this work limits accurate
displacement measurement to time scales of at most 10 seconds or
frequencies of at least 0.1 Hz. Since this is well below 
the lowest eigenfrequencies of the antennas, this is sufficient to
determine dynamic antenna 
behaviour. For the frequency range covered accurately by the
accelerometers, approximately 0.1 to 30 Hz, it is possible to check
the following antenna specifications:

\begin{enumerate}
\item Variations in surface shape, restricted to large scale effects
  like focal length and astigmatism 
\item Variations in pointing in elevation and cross-elevation direction
\item Translation of apex structure with respect to the BUS
\item Variations in path length along the boresight direction
\end{enumerate}

Using detailed long term wind studies, and wind measurements
simultaneous with accelerometer measurements, antenna performance can
be extrapolated to longer time scales under the
assumption that wind effects dominate antenna performance on these
time scales.

Antenna performance should be met for all modes in which the antenna
will be used to perform astronomical observations. For this paper, we
have considered  

\begin{enumerate}
\item pointing where the antenna is commanded to remain fixed at an
  azimuth and elevation position, 
\item sidereal tracking, where azimuth and elevation are updated continuously,
\item fast switching mode, where the antenna is switched quickly
  between two neighbouring points, 
\item interferometric mosaicking, in which areas of sky are mapped at
  slow speed (0.05 deg/s), 
\item on-the-fly mapping (OTF), in which large areas of sky are mapped at
  high scan speed (0.5 deg/s). 
\end{enumerate}

\section{Accelerometer setup}
\label{setup}

\begin{figure}
\resizebox{\hsize}{!}{
%\centering
\includegraphics[scale=0.3]{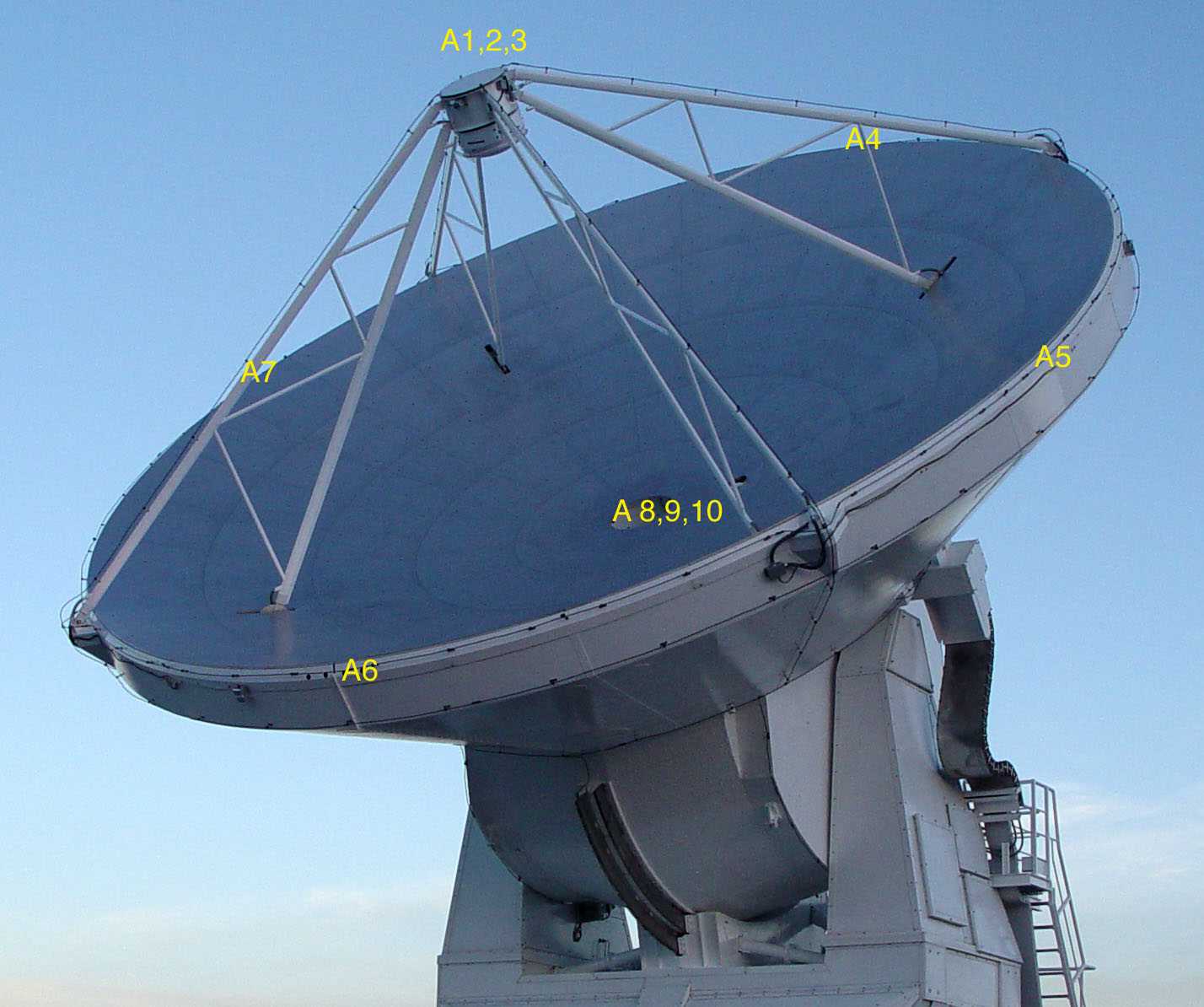}}
\caption{Placement of the accelerometers on the antenna.}
\label{fig:accelconfig}
\end{figure}

We placed 10 accelerometers on the antenna in the following
configuration (Figure~\ref{fig:accelconfig}): 

\begin{itemize}
\item 3 accelerometers as a 3-axis sensor on the subreflector
  structure (A1,2,3)
\item 4 accelerometers along the rim of the BUS in boresight direction (A4-A7)
\item 3 accelerometers as a 3-axis sensor on the receiver flange invar
  ring (A8,9,10)
\end{itemize}

\begin{figure}
\resizebox{\hsize}{!}{
%\centering
\includegraphics[scale=0.40]{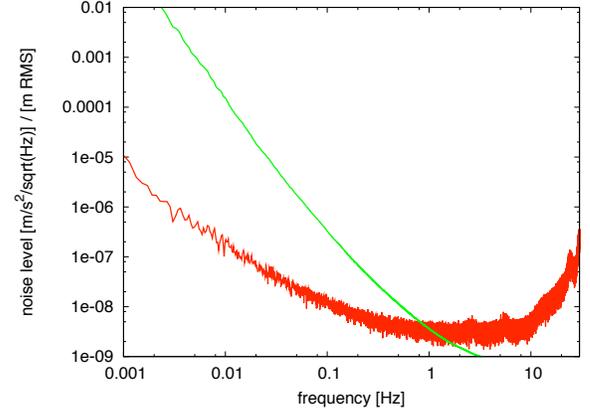}}
\caption{Accelerometer and read-out equipment noise.  The red curve is
the spectral noise in m/s$^2$/$\sqrt(Hz)$, the green curve is the RMS
noise in meters.  The value at frequency $\nu$ Hz shows the typical RMS
noise over timescales of $\frac{1}{\nu}$ seconds.}
\label{fig:accelreadout}
\end{figure}

The accelerometers used for the tests are Endevco model 86 seismic
accelerometers. Together with a multichannel 24 bit A/D
converter, the noise properties are as shown in
Figure~\ref{fig:accelreadout}, red curve. The noise spectrum in
acceleration 
is somewhat constant with frequency in the range 0.1 - 10 Hz, but
since the accelerometers will be used to measure displacement, the
acceleration needs to be integrated twice. This turns the originally
white noise into ``red'' noise, with high power at low frequencies. In
Figure~\ref{fig:accelreadout}, the green curve shows the RMS
displacement noise as a function of frequency, where the integration has
been applied. For frequencies above a few Hz, measurement accuracy is
better than 10 nm, while at 0.1 Hz accuracy has deteriorated to just
below half a $\mu$m. The accelerometers were read out at a frequency of 100
Hz. Accelerometer resonance occurs around 200 - 300 Hz, which required
a hardware anti-alias filter which cuts in above 10 Hz. Effectively
this allows antenna vibrations up to about 30 Hz to be measured. 

On the low frequency side, both the accelerometers and the read-out
electronics cut off frequencies below approximately 0.007 Hz. Note that
the DC or static component can not be measured, which implies that only position
or offset pointing changes can be detected. For most practical
purposes encountered during antenna testing, antenna vibrations below
0.1 to 0.3 Hz are affected by noise. Within the frequency range 0.1 Hz
to 30 Hz the sensitivity for displacements is better than a few
micrometers, or sub-arcsecond for pointing. 

The measurements are time series of accelerations or properties
derived from this. In most cases, this is not particularly
informative, partly because of the limited frequency bandwidth of the
signal. Therefore the choice was made to present most results in the
form of power spectra. Integration of the signal corresponds to
division by the frequency of the power spectrum, and RMS deviations
from the average time series is simply integration over the power
spectrum. In order to reduce noise in the power
spectrum, we can sacrifice low frequency range and frequency
resolution by averaging the power spectra of a number of shorter time
series instead of one long series. Using this noise reduction,
reproducible features show up more clearly in the power spectra. 

The normalisation for the power spectrum used in this work is with the
square of the length $n$ of the time series, and values at negative
frequencies are added to those at positive ones: 

\begin{equation}
\hat{A}(\nu) = \frac{|FFT(A(t))|^2}{n^2}
\label{eq:aoft}
\end{equation}

\noindent{with} $A(t)$ a time series with $n$ samples as a function of
time, $FFT$ the fast Fourier transform, and $\nu$ the frequency. Note
that in the plots in this paper $\sqrt{\hat{A}(\nu)}$ is plotted. The
units along the ordinate reflect this. 

RMS stability of a measured parameter is presented as a function of
frequency:

\begin{equation}
RMS(\nu) = \sqrt{\sum_{\nu'=\nu}^{\infty}{\hat{A}(\nu')}}
\label{eq:rms}
\end{equation}

This type of presentation has the property that at zero frequency, the
RMS is that of $A(t)$, while at higher frequencies the RMS decreases to
include only those contributions from frequencies above and including
$\nu$. An oscillation at a
given frequency shows up as a jump in the RMS curve at this
frequency. In addition to
accelerometer and amplifier noise, thermal effects on the
accelerometers and small tilt variations of the accelerometers in the
gravity field, add additional low frequency noise to the
measurements. 

In order to obtain pointing and displacement data from the
accelerations, the location and orientation of the accelerometers
need to be known. This was determined from known and large antenna
motions, imposed through the drive system. Through combination of
selected accelerometer signals, the following antenna motions could be
isolated:  

\begin{itemize}
\item Elevation pointing (top and bottom accelerometers on the rim of
  the BUS)
\item Cross-elevation pointing (left and right accelerometers on the
  rim of the BUS)
\item Boresight motion of BUS (4 accelerometers on the rim of the BUS,
  one on receiver flange)
\item Path length stability (boresight motion of BUS plus
  accelerometer on apex structure)
\item ``Focal length'' stability or ``defocus''
or ``apex axial motion'' of BUS (4
  accelerometers on the rim of the BUS, one on receiver flange)
\item ``+'' astigmatism of BUS (4 accelerometers on the rim of the
  BUS)
\end{itemize}

The accelerometers on the BUS are configured to measure deformations
which can be described with low order Zernike
polynomials \cite{Zernike1934}. Path length changes, caused by
motion of the entire BUS 
as a rigid body along the boresight direction, can be described as the
(n=0, m=0) $Z_0^0$ ``piston'' term; changes in elevation pointing as
the (n=1, m=1) $Z_1^1$ ``tilt'' term; changes in cross-elevation
pointing as the (n=1, m=-1) $Z_1^{-1}$ ``tip'' term; boresight motion of
the rim of the BUS with respect to the receiver flange, considered to
be representative for the vertex of the antenna, as the (n=2, m=0)
$Z_2^0$ ``defocus'' term; boresight deformation of the left and right
side of the rim of the BUS in opposite direction as the top and bottom
side of the BUS, as the (n=2, m=2) $Z_2^2$ ``+ - astigmatism''
term. Due to the limited number of accelerometers on the rim of the
BUS, the (n=2, m=-2) $Z_2^{-2}$ ``$\times$ - astigmatism'' term can not be
measured.

\section{Test Conditions}

Environmental conditions have a significant influence on the
performance of the antenna. On the time scales relevant for the
accelerometer measurements, the major environmental effects are caused
by the wind. To draw meaningful conclusions from the measurements, it
was thus necessary to carefully characterise the local wind through
wind measurements both independent of as well as simultaneous with the
accelerometer measurements. 

In addition to variable wind conditions, the thermal environment
affects antenna performance. Since thermal effects of the antenna are
on timescales longer than those accurately measurable with the
accelerometers, they will not be considered for the measurements
presented here \cite{Greve2006}.

\subsection{Wind Conditions}

Wind data at the ATF were recorded over a period of more than a year,
sampled continuously at 10 Hz with a sonic anemometer, placed some 30
m north of the antennas. The power spectrum of the wind speed over
intervals of approximately 1000 seconds was determined as a function of average wind
speed and direction. 

The local terrain and building geometry affect the wind power
spectrum. Overall, the terrain is reasonably flat out to more than 10
km in any direction. Seen from the anemometer, the VLA control
building is located a few hundred meters towards the west, the
Mitsubishi prototype antenna about 30 m to the south-west, the
VertexRSI prototype antenna about 30 m to the south-south-east, and
the AEC prototype antenna about 50 m to the south-east.  

The wind power spectrum has a typical slope of $\nu^{-1.5}$, where $\nu$ is the
frequency. From theory, an exponent of $-\frac{5}{3}$ is expected for the
microscale range, while an exponent of -1 is expected for the mesoscale
turbulence range. The measured exponent is consistent with theory,
showing a transition between micro- and mesoscale turbulence. 

The effect of obstructions on the power spectrum seems to increase the
high frequency power in the spectrum, while keeping the same
exponent. In addition, the low frequency part is lowered and the
exponent is decreased towards lower frequencies. The observed
transition between low and high frequency is around 0.1 Hz. 

\begin{figure}
\resizebox{\hsize}{!}{
%\centering
\includegraphics[scale=0.4]{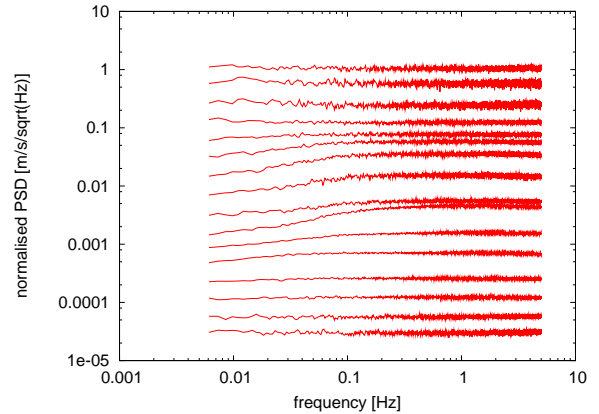}}
\caption{Wind power spectra, normalised with the typical undisturbed
  wind spectrum. The 16 curves show equally spaced wind directions,
  centered on the sonic anemometer, starting at north (top) and
  increasing clockwise. Each subsequent spectrum is shifted by a
  factor 0.5 for clarity. Only wind speeds over 2.5 m/s were used.}
\label{fig:windpowerspectra}
\end{figure}

Figure~\ref{fig:windpowerspectra} shows the wind power spectra,
normalised with the typical undisturbed wind spectrum, for 16 equally
spaced wind directions, starting at the north and stepping through the
east. Only wind speeds over 2.5 m/s were used.  

\begin{figure}
\resizebox{\hsize}{!}{
%\centering
\includegraphics[scale=0.4]{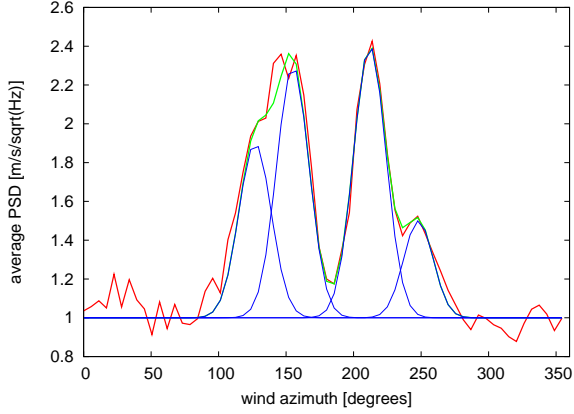}}
\caption{Normalised wind power spectra averaged over 0.3 - 2.0 Hz, as
  a function of wind direction, as seen from the sonic anemometer. The
  peak between 90 and 180 degrees is the combined effect of the AEC
  and VertexRSI antenna, the peak between 180 and 240 is due to the
  Mitsubishi antenna, and the peak around 250 degrees is due to the
  VLA control building.  The red curve shows the measured turbulence
  component, while the green curve shows an analytical approximation
  composed of 4 gaussian components matched to the observed peaks. The
  4 individual components are shown as blue lines.}
\label{fig:windpowernorm}
\end{figure}

Figure~\ref{fig:windpowernorm} shows the normalised power spectra of
Figure~\ref{fig:windpowerspectra} averaged over the frequency range
between 0.3 and 2.0 Hz, plotted as a function of azimuth. The shape of
the curve can easily be explained by the positions of the antennas and
the VLA control building.

The wind power spectra show distinct features depending on the wind
direction. The sonic anemometer can clearly distinguish the VLA control
building, Mitsubishi antenna, the clearance between the Mitsubishi and
VertexRSI antennas, and the combined effect of the VertexRSI and AEC
antennas. This directional dependence complicates interpretation of
the wind results.  Because any obstruction will be at different
azimuths for the antenna and the sonic anemometer,  they will be
subjected to a different wind spectrum. 

Three of the four peaks in Figure~\ref{fig:windpowernorm} show the
wake turbulence of an antenna. This information is used to correct the
wind spectra as experienced by the antennas in order to predict performance
for undisturbed wind conditions, but can also be used to predict
performance for an antenna in the compact configuration of the ALMA
array, where the antennas are only a few tens of meters apart. 

The wind power spectra are needed to scale the antenna vibration power
spectra to that for a reference wind spectrum. Since for short
(typically 15 minutes) measurements the error bars on a single power
spectrum can be large, and the natural variations in the wind can
throw off an observed power spectrum far away from the average, the
following wind spectral features were used to treat wind effects. 

\begin{enumerate}
\item Effects of upwind obstructions change the shape of the power
  spectrum, but above frequencies of approximately 0.2 Hz the result
  is effectively a change in the level of the power spectrum, while
  the slope remains unchanged and close to the theoretically expected
  value (Figure~\ref{fig:windpowerspectra}). 
\item For frequencies above approximately 1 Hz, effects of aliasing
  become visible, introducing errors in the measured slope. 
\end{enumerate}

These effects combined leave a frequency interval between 0.2 and 1.0
Hz where the slope of the power spectrum is reasonably well-defined,
and independent of upwind obstructions. This interval is used for
normalisation of the power spectra, instead of the average wind speed,
which may be affected by obstructions and stochastic effects.  

Prediction of low frequency wind effects was done using the average
power spectra calculated from all available wind data, and
extrapolation from the 0.2 - 1.0 Hz range. Known obstructions
(Figure~\ref{fig:windpowernorm}, and a modification of it to reflect
the obstructions as seen by the antennas) and variation of the power
spectra with wind speed were taken into account.

\subsection{Wind Scaling}
\label{windscaling}

The dynamic wind pressure on the structure is a function of
frequency. For a single point, the pressure can be directly coupled to
the power spectrum of the square of the wind speed. For extended
structures spatial averaging will occur. Thus, for turbulence of a
geometric scale smaller than the typical dimension of the structure,
the effective pressure is less than for turbulence of larger scale,
given the same power spectral density: the antenna serves as a
mechanical low-pass filter. The averaging effect of the dynamic wind
pressure can be described with the aerodynamic admittance function
(AAF) of the form 

\begin{equation}
AAF(\nu) = \frac{1}{1+\left(2\nu\frac{L}{U}\right)^{\frac{4}{3}}}
\label{eq:aaf}
\end{equation}

\noindent{where} $L$ is the typical scale of the
structure, and $U$ is the average wind speed
\cite{Davenport1961}. Since the average wind speed 
varies from measurement to measurement, the AAF also varies.

The (static) wind force on a structure is given by

\begin{equation}
F_{wind}= \frac{1}{2}\cdot C_d \cdot L^2 \cdot \rho \cdot v^2
\label{eq:stat_wind}
\end{equation}

\noindent{where} $C_d$ is the aerodynamical drag coefficient, $\rho$ the air density, and $v$ the
wind speed. Since the antennas are fixed to the ground, the wind force
will bend the structure, where the stiffness $k$ of the structure
determines the amount of flexure $x$: 

\begin{equation}
F_{wind}= - k \cdot x
\label{eq:stiffness}
\end{equation}

Combination of equations \ref{eq:stat_wind} and \ref{eq:stiffness} yields:

\begin{equation}
x = - \frac{C_d \cdot L^2 \cdot \rho \cdot v^2 }{2 \cdot k}
\end{equation}

This is assumed to be valid for any antenna deformation or motion, and
to hold for any frequency, and thus for the entrire power spectral density
(PSD) curve: 

\begin{equation}
PSD_{antenna}(\nu) = PSD_{wind speed}(\nu) \cdot H(\nu) \cdot AAF(\nu)
\cdot \rho
\label{eq:psd}
\end{equation}

\noindent{which} defines $H(\nu)$ as the transfer function between wind power spectrum
and antenna motion power spectrum. The drag coefficient $C_d$ as well as the effective
area $L^2$ of the antenna, and thus the transfer function $H(\nu)$, depend on the
orientation of the antenna in azimuth and elevation relative to the wind.

\section{Analysis Methods}
\label{methods}

The accelerometer sensitivity was calibrated against gravity around
0.1 Hz, using the antenna to tilt the accelerometers in a controlled
way. The sensitivity and frequency response of the read-out
electronics was calibrated using a switchable resistor, with which a
step-function was fed to the electronics. All relevant measurement
data were corrected for accelerometer and electronic effects. 

Signals were combined in the time domain, after which the power
spectra were calculated. Frequency response corrections were done in
the frequency domain. Double integration of the accelerometer signals,
needed to obtain displacement and pointing information, was achieved
through division in the frequency domain by $(2 \pi \nu ) ^2$.
See the Appendix for details.

\subsection{Extrapolation to Lower Frequencies}

The accelerometers are limited in the lowest frequency at which
displacements can be measured accurately. Depending on the size of the
displacements measured by the accelerometers, and canceling of
(tilt-)noise terms in the analysis, the lowest frequency varies
between 0.08 and 1 Hz. This frequency limit is well below the lowest
eigenfrequency of the antennas, which are 6-7 Hz. 

From dynamic structure behaviour, a constant stiffness is expected at
frequencies well below the lowest eigenfrequency. This means that the
dynamic stiffness of the antenna should reach a constant value in the
frequency range between the accelerometer low noise limit and the
lowest eigenfrequency. This would show in the plots of the transfer
functions H as a flattening of the curve towards lower
frequencies. This flattening is seen in all measurements of
wind-excited motion. Thus it is possible to determine the (constant)
value of the low frequency stiffness from the flat part in the
transfer functions, and extrapolate this constant value to lower
frequencies, replacing the values in the transfer function affected by
accelerometer noise. The scatter in the flat section of the transfer
function is used to determine the extrapolation uncertainty, which
typically is 6-9\%. Subsequently, the extrapolated transfer function
is used to calculate antenna performance for the assumed or measured
wind spectrum, and the aerodynamic admittance function for 9 m/s
wind, as required by the specifications.

\subsection{Pointing Accuracy}

The simplest operational mode for the antenna is pointing to a fixed
azimuth and elevation, whereby the drives are powered and the brakes
are released. In this mode wind effects are best investigated, since
any effects of antenna shake due to position updates from the control
system are minimal. During sidereal tracking the antenna drives are
constantly updating the azimuth and elevation positions, including the
azimuth and elevation speed, in order to achieve a smooth 
tracking motion. Updates are performed at 48 ms intervals,
corresponding to a frequency of 20.83 Hz. Antenna motion at
frequencies below half this value can be controlled by the drive
system, while higher frequencies can only be excited by the drives but
not actively corrected. 

Wind-induced pointing jitter was investigated for 11 wind-oriented
azimuth/elevation combinations, and sidereal tracking was checked for
30 azimuth/elevation combinations spread evenly over the sky. Calm
wind conditions were chosen for the sidereal tracking measurements, in
order to clearly separate effects due to the drive system from those
due to wind. The total pointing accuracy for the antenna is the
squared sum of the pointing for high wind conditions without tracking,
and the pointing for sidereal tracking during low wind conditions. 

The 4 accelerometers on the rim of the BUS were used for deriving the
BUS pointing accuracy. The positions for the 4 accelerometers are at
the top, right, bottom and left of the BUS, when looking into the
antenna at low elevation. Differential signals between two
accelerometers were used to eliminate the large gravity signal common
to all accelerometers and being a function of elevation. The
cross-elevation and elevation pointing $\Delta_{XEL}$ and
$\Delta_{EL}$ are calculated as follows: 

\begin{eqnarray}
\alpha_{XEL} &=& \frac{a_l - a_r}{x_l - x_r} \\
\alpha_{EL} &=& \frac{a_t - a_b}{y_t - y_b} \\
\Delta_{XEL} &=& \int\int \alpha_{XEL} d^2t \\
\Delta_{EL} &=& \int\int \alpha_{EL} d^2t
\label{eq:pointing1}
\end{eqnarray}

\noindent{where} $\alpha_{XEL}$ is the cross-elevation angular
acceleration, $\alpha_{EL}$ is the elevation angular acceleration,
and $a_t$, $a_r$, $a_l$, and $a_b$ are the top, right, left, and
bottom accelerometer signals, respectively.  The variables x and y are
the positions of the accelerometers in coordinates along the elevation
axis and perpendicular to the elevation axis, respectively, with the
assumed intersection of elevation axis and azimuth axis as the origin.
The integration is over the time coordinate t.  The total pointing
error is given by:

\begin{equation}
\Delta_{TOT} = \sqrt{\Delta^2_{XEL} + \Delta^2_{EL}}
\label{eq:pointing2}
\end{equation}

\noindent{which} assumes uncorrelated pointing jitter in elevation and
cross-elevation. The accelerometer signals are not significantly
affected by noise and other errors down to a frequency of about 0.1 -
0.2 Hz. Below this frequency, the apparent pointing error increases
faster than what can be expected from the wind power spectrum and
conservative assumptions on the dynamic behaviour of the antenna
structure. In combination with optical pointing telescope (OPT) or
radiometer measurements, it is possible to measure pointing behaviour
for frequencies lower than 0.1 Hz (see Figure~\ref{fig:OPT_wind_stiffness}) . 

The transfer functions show a flat section below the lowest
eigenfrequency. As explained above, this is expected from dynamic
antenna behaviour and allows extrapolation of the antenna properties
to lower frequencies, as well as the use of the OPT which follows
the bulk motion of the reflector in this frequency range. This has
been done in the corresponding plots,
where the specified wind spectrum has been used to predict antenna
properties for time periods up to 15 minutes (0.001 Hz). 

Encoder data recorded simultaneously with the wind and accelerometer
data were used to calculate the encoder errors, defined as the measured
encoder read-out minus the commanded antenna position. Azimuth encoder
errors were converted to cross-elevation errors through multiplication
with the cosine of the elevation.

\subsection{Primary Reflector Deformation}

The combined accelerometer signals measured at the rim of the BUS and at
the receiver flange give information about some low-order deformations of
the BUS, and thus about the accuracy of the primary reflector surface. 
Numbers for the deformations are expressed as boresight components of the deformation
at the location of the rim of the BUS. No averaging over the entire
reflector surface is attempted. BUS ``+'' astigmatism is defined as
follows: 

\begin{equation}
Astig \equiv \int\int a_t + a_b - a_r - a_l d^2t
\label{eq:astig}
\end{equation}

\noindent{and} apex axial motion ($AAM$) is given by:

\begin{equation}
AAM \equiv \int\int \frac{\left(a_t + a_r + a_b + a_l\right)}{4} -
a_{rf} d^2t
\label{eq:focus}
\end{equation}

\noindent{where} $a_{rf}$ is the accelerometer signal measured on the
receiver flange and in boresight direction.

\subsection{Path length variations}

Path length changes measured with the accelerometers reflect the
boresight motion of the BUS in the inertial coordinate system, plus
the effects of distance changes between the apex and receiver flange: 

\begin{eqnarray}
\omega_{xel} &=& \int \alpha_{xel} dt \\
\omega_{el} &=& \int \alpha_{el} dt \\
\alpha_{cf} &=& -\left(\omega^2_{xel} + \omega^2_{el}\right) \cdot Z_{apex} \\
a_{BUS} &=& \frac{a_t + a_r + a_b + a_l + 4a_{rf}}{8} \\
pl &=& \int\int 2\left(a_{apex} - \alpha_{cf}\right) - a_{BUS} d^2t
\label{eq:pathlength}
\end{eqnarray}

where $Z_{apex}$ is the distance between the apex boresight
accelerometer and the azimuth and elevation axes, $a_{apex}$ is the
boresight acceleration component at the apex, and $\alpha_{cf}$ is the
centrifugal acceleration at the apex due to elevation and
cross-elevation motion. The path length calculated this way represents
the total optical path length assuming rigid body motion of the BUS
and receiver flange, and allowing for boresight displacements of the
apex structure. Since the accelerometers measure in the inertial
system, and are aligned with the antenna boresight, there is no need
to refer to the ground coordinate system or measure mount and yoke
path length variations.

\subsection{Structural flexure}
%================
\label{flexure}

Combination of encoder and accelerometer pointing information allows
for calculation of antenna structure deformation.  The total
deformation of the structure between the encoders and the
accelerometers on the rim of the BUS was determined by integrating the
angular accelerations measured at the BUS to obtain a time series of
the angle. Small corrections for timing differences between the
encoder and accelerometer equipment were applied.
Encoder read-out (AZ$_{enc}$ and EL$_{enc}$, with the sine and cosine of
EL$_{enc}$, $\sin_{el}$ and $\cos_{el}$) and the local acceleration
$g$ due to gravity are used to calculate the
expected accelerations on any point of the antenna with coordinates
x,y,z with the intersection of the azimuth and elevation axes as
origin, the coordinate system is fixed to the reflector, with x along
the elevation axis, y upward and z along the optical axis for elevation=0,
with the axis of measurement of an accelerometer pointing in the
direction $\overrightarrow{dir}$:

\begin{eqnarray}
\omega_{enc,az} &=& \frac{d AZ_{enc}}{dt} \\
\omega_{enc,el} &=& \frac{d EL_{enc}}{dt} \\
\alpha_{enc,az} &=& \frac{d \omega_{enc,az}}{dt} \\
\alpha_{enc,el} &=& \frac{d \omega_{enc,el}}{dt} 
\end{eqnarray}
\begin{eqnarray}
\overrightarrow{g} &=& g \cdot 
\begin{pmatrix}
0 \\
-\cos_{el} \\
-\sin_{el}
\end{pmatrix} 
\end{eqnarray}
\begin{eqnarray}
\overrightarrow{cf_{el}} &=& \omega^2_{enc,el} \cdot 
\begin{pmatrix}
0 \\
y \\
z
\end{pmatrix} 
\end{eqnarray}
\begin{eqnarray}
\overrightarrow{cf_{az}} &=& \omega^2_{enc,az} \cdot 
\begin{pmatrix}
x \\
y \cdot \sin^2_{el} - z \cdot \sin_{el} \cdot \cos_{el} \\
-y \cdot \sin_{el}\cos_{el} + z \cdot \cos^2_{el}
\end{pmatrix} 
\end{eqnarray}
\begin{eqnarray}
\overrightarrow{ang_{el}} &=& \alpha_{el} \cdot 
\begin{pmatrix}
0 \\
z \\
y
\end{pmatrix} 
\end{eqnarray}
\begin{eqnarray}
\overrightarrow{ang_{az}} &=& \alpha_{az} \cdot 
\begin{pmatrix}
y \cdot \sin_{el} - z \cdot \cos_{el} \\
x \cdot \sin_{el} \\
x \cdot \cos_{el}
\end{pmatrix} 
\end{eqnarray}
\begin{eqnarray}
\overrightarrow{accel} &=& \overrightarrow{g} + \overrightarrow{cf_{el}} + \overrightarrow{cf_{az}} +
\overrightarrow{ang_{el}} + \overrightarrow{ang_{az}} 
\end{eqnarray}
\begin{eqnarray}
sig_{accel} &=& \overrightarrow{accel} \cdot
\begin{pmatrix}
dir_x \\
dir_y \\
dir_z
\end{pmatrix}
\label{eq:flexure}
\end{eqnarray}

The centrifugal accelerations due to elevation and azimuth slews are given
by $\overrightarrow{cf_{el}}$ and $\overrightarrow{cf_{az}}$ respectively,
and the accelerations due to angular accelerations in elevation and
azimuth are given by $\overrightarrow{ang_{el}}$ and $\overrightarrow{ang_{az}}$
respectively.
The expected acceleration thus calculated is then convolved with the
time response $TR$ of the read-out equipment, to obtain a time series
of expected accelerations which can be directly compared to the
measured accelerations:

\begin{equation}
\alpha_{expected} = sig_{accel} \ast TR
\label{eq:accelexp}
\end{equation}

\noindent{where} $\ast$ denotes convolution, and $\alpha_{expected}$
is the expected bandwidth filtered accelerometer signal.

Both measured and expected accelerometer signals are treated in the
same way to calculate antenna pointing. The data are resampled to a
common time grid, small corrections for accelerometer sensitivity and
integration constants are applied to make the expected and measured
antenna pointing match, and the difference between the measured and
expected pointing is plotted together with a scaled curve of the
angular acceleration during the fast motion of the antenna.
The scaling factor required to make the acceleration curve match to
the structure flexure curve is the measure of flexure of the antenna
structure.
Since the flexure due to an inertial acceleration scales with the
acceleration and the mass of the antenna, and gravitational flexure of
the antenna also scales with the mass of the antenna, the numbers for
the structure flexure give an indication of generic structure
stiffness against gravitational effects as well.

\section{Results}

The measurement results obtained at the ATF site were used to derive
environmentally independent antenna properties where possible. The
Statement of Work and Specifications for the ALMA prototype antennas (SoW)
specifies antenna performance for the environmental conditions at the
ALMA site, where in particular the wind power spectrum and air density
are of relevance for this investigation. Using the transfer funcions
derived at the ATF site, antenna performance was calculated for
conditions representative at the ALMA site and defined in the SoW.
The main differences between the ATF and SoW conditions are for air pressure
(SoW: 550 mBar), wind speed (ATF: variable between 0.05 m/s and exceeding 
20 m/s, SoW: 9 m/s) and wind power
spectrum (SoW: more power at lower frequencies than observed at ATF).

\subsection{Pointing}

Table \ref{tab:accelpoint} summarises the dynamic pointing performance
of the AEC and VertexRSI antennas. Measurement noise is typically
sub-$\mu$m scale at 0.1 Hz, and several nm at 1 Hz. Uncertainties in
the results are generated by the variation of the measurement results
over different elevation and azimuth positions, uncertainties in the
determination of the wind power spectrum as experienced by the antennas
at the time of measurement, and extrapolation to lower frequencies.
Where applicable, these three individual contributions to the measurement
results are shown in Table~\ref{tab:accelpoint}.

\begin{table*}
\centering
\caption{Pointing}
\begin{tabular}{|l|c|c|}
\hline
Pointing type & VertexRSI & AEC \\
\hline
stationary, windy conditions & 
$0.81~\pm0.24^a~\pm0.05^b~\pm0.20^c$ arcsec&
$0.45~\pm0.10^a~\pm0.02^b~\pm0.11^c$ arcsec \\

sidereal tracking, no wind &
$0.47~\pm~0.11^a$ arcsec &
$0.22~\pm~0.08^a$ arcsec \\

tracking, windy conditions, 0.1 Hz &
$0.58~\pm~0.15^a~\pm~0.08^c$ arcsec &
$0.29~\pm~0.09^a~\pm~0.05^c$ arcsec \\

tracking, windy conditions, 0.001 Hz &
$0.94~\pm~0.26^a~\pm~0.05^b~\pm~0.20^c$ arcsec &
$0.50~\pm~0.13^a~\pm~0.02^b~\pm~0.11^c$ arcsec \\

on-the-fly, 0.5 deg/s, 1 Hz &
$1.7~\pm~0.7$ arcsec &
$0.5~\pm~0.3$ arcsec \\

on-the-fly$^d$, 0.5 deg/s, 1 Hz &
&
$0.372~\pm~0.015$ arcsec \\

on-the-fly, 0.05 deg/s, 1 Hz &
$0.8~\pm~0.5$ arcsec &
$0.231~\pm~0.007$ arcsec \\

\hline
\multicolumn{3}{l}{$^a$~Spread over different azimuth/elevation
  combinations.} \\
\multicolumn{3}{l}{$^b$~Extrapolation error for the transfer
  function.} \\
\multicolumn{3}{l}{$^c$~Uncertainty in wind power spectrum
  determination.} \\
\multicolumn{3}{l}{$^d$~Ignoring AEC apex rotation.} \\
\end{tabular}
\label{tab:accelpoint}
\end{table*}

\subsubsection{Stationary pointing}

\begin{figure}
\resizebox{\hsize}{!}{
%\centering
\includegraphics[scale=0.35]{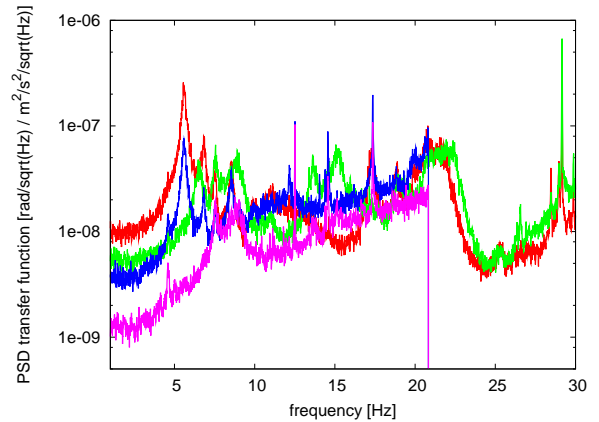}}
\caption{VertexRSI antenna elevation (red) and cross-elevation (green)
  pointing transfer functions for wind excitation. The blue and magenta
  curves show the encoder error transfer functions for elevation and
  cross-elevation respectively.}
\label{fig:vertex-pttfunc}
\end{figure}

Figure~\ref{fig:vertex-pttfunc} shows the transfer functions H between
wind PSD and BUS pointing PSD for the VertexRSI antenna, as defined in 
\S\ref{windscaling}. The
red and green curves show elevation and cross-elevation pointing
respectively. The low-frequency values of the red and green curves,
here shown for frequencies above 1 Hz, show the expected behaviour
below the lowest eigenfrequency and can be extrapolated to 0
Hz. Similar curves and behaviour are observed for the AEC antenna,
though not shown here. 

Wind-induced pointing jitter is dominated for both antennas by
elevation motion.
Encoder errors do not exceed 0.14
arcsec for the VertexRSI antenna, and remain below 0.07 arcsec for the AEC antenna.
Both antennas show azimuth-dependendent pointing performance, where
pointing jitter is larger when the antenna is looking into the wind,
and smallest for wind coming from sideways-behind. 
These results most likely reflect the smaller projected area of the antenna
when viewed from the side, and the higher drag coefficient of the ``cup''
formed by the primary mirror and BUS when viewed from the front.

\subsubsection{Sidereal Tracking}

Tracking jitter of the VertexRSI antenna is for a significant part due
to elevation motion, with large contribution in the 3-6 Hz
range. Total tracking jitter over timescales of 10 seconds amounts to
$0.47~\pm~ 0.11$ (spread) arcsec average. Largest jitter is
observed for low elevation in the south-east and south-west, and
minimum tracking jitter is seen while crossing the meridian. Total
tracking jitter for the AEC antenna, over timescales of 10 seconds,
amounts to $0.22~\pm~0.08$ (spread) arcsec average. As for
the VertexRSI antenna, the sidereal tracking jitter depends on the
antenna pointing. 

\begin{figure}
\resizebox{\hsize}{!}{
%\centering
\includegraphics[scale=0.35]{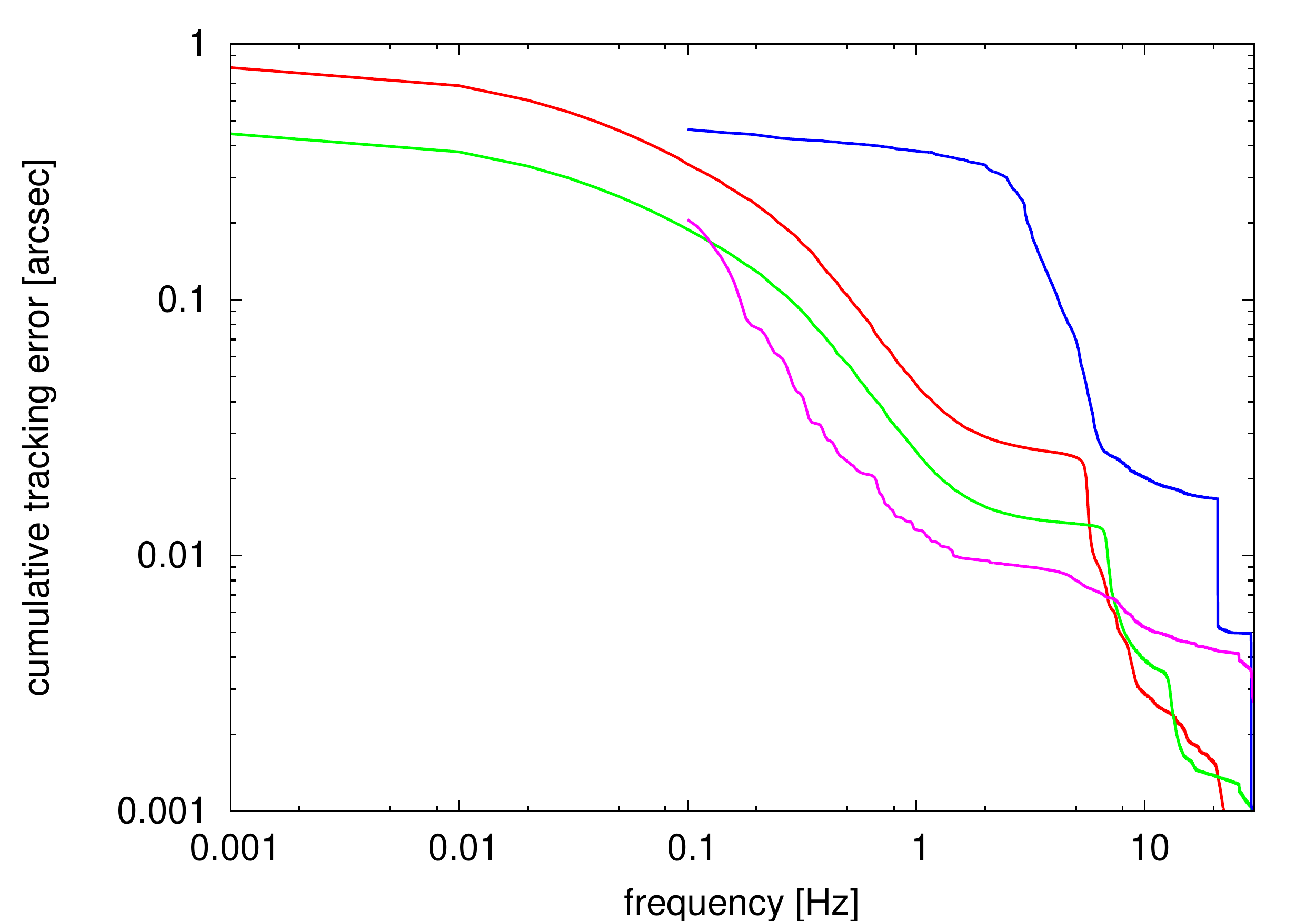}}
\caption{RMS tracking jitter for VertexRSI and AEC antennas. The
  extrapolated wind-induced jitter for conditions specified in the SoW
  is shown as red (VertexRSI) and green (AEC) curves. Pointing
  stability during sidereal tracking is given by the blue (VertexRSI)
  and magenta (AEC) curves. For sidereal tracking under windy conditions,
  both curves must be combined.}
\label{fig:trackPSD}
\end{figure}

\subsubsection{Combined Wind and Tracking}

The pointing jitter due to wind and due to sidereal tracking are
assumed to be uncorrelated, in which case the cumulative pointing
error curves can be added quadratically. A check of an independent
measurement with high wind and sidereal tracking showed that this
assumption is valid. Since no tracking jitter at 0.001 Hz could be
measured, a conservative value equal to the observed tracking jitter
at 0.1 Hz is used.  

Figure~\ref{fig:trackPSD} shows the wind induced and sidereal tracking
jitter for both antennas. At 0.1 Hz, the combined wind and tracking
jitter was calculated from direct measurements, while at 0.001 Hz the 
pointing jitter was derived from the extrapolated transfer function.
The red and green curves show the wind-induced pointing jitter for
the VertexRSI and AEC antennas respectively, and the blue and magenta
curves show the sidereal tracking stability.

\subsubsection{Pointing During OTF}

During OTF scanning the time between switches in scan direction
was typically 10 seconds. With a few seconds settling time for
the antenna after each switch, only a few seconds remained during
which representative measurements could be taken. This limited the
effective lowest frequency at which antenna performance could be 
calculated to 1 Hz.

For the VertexRSI antenna, the azimuth and elevation at which the scan
was performed had large impact on the pointing stability during the scan.
On the other hand, the AEC antenna measurements were affected
for some parts of the scan by apex rotation feeding back to the BUS
motion, an effect which died out about a minute into the scan. 
Both effects resulted in a large spread of the measurement results,
represented in the standard deviation stated in Table~\ref{tab:accelpoint}.

\subsection{BUS deformations}

\begin{table*}
\centering
\caption{BUS Deformation}
\begin{tabular}{|l|c|c|}
\hline
Type of deformation & VertexRSI & AEC \\
\hline
astigmatism, wind-induced, 0.001 Hz & 5.3~$\mu$m & 6~$\mu$m \\
astigmatism, 1.5 - 10 s after fast-switch & 2~$\mu$m peak-to-peak & 15~$\mu$m peak-to-peak \\
astigmatism, OTF 0.5 deg/s, 1 Hz & $0.9~\pm~0.3$~$\mu$m & $4~\pm~2$~$\mu$m \\
astigmatism, OTF$^a$ 0.5 deg/s, 1 Hz & & $3.19~\pm~0.09$~$\mu$m \\
astigmatism, OTF 0.05 deg/s, 1 Hz & $0.27~\pm~0.09$~$\mu$m & $2.2~\pm~0.9$~$\mu$m \\
%defocus
AAM, wind-induced, 0.001 Hz & 2.2~$\mu$m & 5~$\mu$m \\
%defocus
AAM, OTF, 0.5 deg/s, 1 Hz & $1.9~\pm~0.6$~$\mu$m & $2 - 20$~$\mu$m \\
%defocus
AAM, OTF$^a$, 0.5 deg/s, 1 Hz &  & $1.94~\pm~0.08$~$\mu$m \\
%defocus
AAM, OTF, 0.05 deg/s, 1 Hz & $1.0~\pm~0.3$~$\mu$m & $1.4~\pm~0.2$~$\mu$m \\
%defocus
AAM and astigmatism, tracking induced, 0.1 Hz & $< 1$~$\mu$m & $< 1$~$\mu$m \\

\hline
\multicolumn{3}{l}{$^a$~Ignoring AEC apex rotation.} \\
\end{tabular}
\label{tab:busdeform}
\end{table*}

Table~\ref{tab:busdeform} summarises the results for deformation of the BUS.
For both antennas, surface stability is dominated by the stiffness of the
BUS for wind excitation, and apex axial motion %defocus
 and astigmatism are each below 1
$\mu$m RMS for sidereal tracking.

During a fast switch of antenna pointing, the accelerometers show a
deformation of the BUS rim of nearly 1.4 mm astigmatism peak-to-peak
for the VertexRSI antenna, and nearly 4 mm for the AEC antenna.
During a fast switch, no astronomical data is recorded, so deformation
of the BUS is no concern, as long as recovery of the shape is fast enough
after the antenna pointing has stabilised.
Astigmatism recorded 1.5 seconds after the fast switch started, up to
the next switch 8.5 seconds later, remains well below 2 $\mu$m
peak-to-peak for the VertexRSI antenna, and below 15 $\mu$m
peak-to-peak for the AEC antenna, after removal of a low order
polynomial. The polynomial removes the large noise component in the
accelerometer signals at the lowest frequencies. In this case, the
crude removal of the noise component is justified, but the resulting
number for the remaining astigmatism should only be used as an order
of magnitude indication. 

Five seconds after a fast switch for the AEC antenna, the astigmatism
had died down to typically 1 $\mu$m peak-to-peak. The reason for the
large peak-to-peak variation is the 5 Hz resonance of the apex
structure which takes some time to die out. During apex rotation, the
BUS gets deformed through the feed legs bending which drive the apex
rotation.  

BUS astigmatism during the fast OTF scan, with 0.5 deg/s scan rate,
averages to $0.9~\pm~0.3$ $\mu$m RMS over timescales of 1 second for
the VertexRSI antenna, and $4~\pm~2$ $\mu$m RMS for the AEC
antenna. Surface stability is affected for some parts of the scan by
apex rotation feeding back to the BUS motion. The spread of 2 $\mu$m
($1~\sigma$) reflects this variable
surface stability. Ignoring the parts of the scan affected by apex
rotation, the numbers reduce to $3.19~\pm~0.14$ $\mu$m. When the scan
rate is reduced to 0.05 deg/s, for interferometric mosaicking, BUS
astigmatism averages to $0.27~\pm~0.09$ $\mu$m RMS over timescales of
1 second for the VertexRSI antenna, and $2.2~\pm~0.2$ $\mu$m RMS for
the AEC antenna.  

BUS %defocus
apex axial motion stability during the fast OTF scan, with 0.5 deg/s scan
rate, averages to $1.9~\pm~0.6$ $\mu$m RMS over timescales of 1 second
for the VertexRSI antenna, and 2 to 20 $\mu$m RMS for the AEC
antenna. Also here, the apex rotation significantly affected the
surface stability of the BUS. Ignoring the parts of the scan affected
by apex rotation, the numbers reduce to $1.94~\pm~0.08$ $\mu$m. When
the scan rate is reduced to 0.05 deg/s, for interferometric
mosaicking, VertexRSI BUS %defocus
AAM stability averages to $1.0~\pm~0.3$
$\mu$m RMS over timescales of 1 second, and AEC BUS %defocus
AAM stability
averages to $1.4~\pm~0.2$ $\mu$m RMS. 

Overall, dynamic BUS deformations due to motion of the antennas are 
small if the apex rotation effect can be ignored.

\subsection{Path Length}

\begin{table*}
\centering
\caption{Path Length}
\begin{tabular}{|l|c|c|}
\hline
Source of path length variations & VertexRSI & AEC \\
\hline
wind-induced, 0.001 Hz & 6~$\mu$m & 6~$\mu$m \\
sidereal tracking, 1 Hz & 2~$\mu$m & 0.5~$\mu$m \\
OTF, 0.5 deg/s, 1 Hz & $12~\pm~7$~$\mu$m & $3.3~\pm~2.9$~$\mu$m \\
OTF$^a$, 0.5 deg/s, 1 Hz&  & $2.2~\pm~0.5$~$\mu$m \\
OTF, 0.05 deg/s, 1 Hz & $3.1~\pm~1.0$~$\mu$m & $0.7~\pm~0.02^b$~$\mu$m \\

\hline
\multicolumn{3}{l}{$^a$~Ignoring AEC apex rotation.} \\
\multicolumn{3}{l}{$^b$~BUS only.} \\
\end{tabular}
\label{tab:pathlength}
\end{table*}

Wind induced path length variations for both antennas amount to 6
$\mu$m RMS over timescales of 15 minutes, see Table \ref{tab:pathlength}
for an overview

For sidereal tracking, path length variations over timescales of 1
second remain below 2 $\mu$m for the VertexRSI antenna, and below 0.5
$\mu$m for the AEC antenna. 

Total path length stability during the fast OTF scan, with 0.5 deg/s
scan rate, averages for the VertexRSI antenna to $12~\pm~7$ $\mu$m RMS
over timescales of 1 second. Depending on the azimuth, path length
stability may be as bad as 20 $\mu$m RMS. The AEC antenna path length
jitter averages to $3.3~\pm~2.9$ $\mu$m RMS. The path length stability
is affected for some parts of the scan by apex rotation feeding back
to the BUS motion, as well as changing the distance between BUS and
apex. The spread of 2.9 $\mu$m
($1\sigma$) reflects this variable path length stability. Ignoring the
parts of the scan which are affected by apex rotation, the numbers
reduce to $2.2~\pm~0.5$ $\mu$m.  

When the scan rate is reduced to 0.05 deg/s, for VertexRSI antenna
interferometric mosaicking, total path length stability averages to
$3.1~\pm~1.0$ $\mu$m RMS over timescales of 1 second, and
$0.70~\pm~0.02$ $\mu$m RMS for the AEC antenna. Note that for the AEC
antenna, this number covers the BUS boresight path length stability
only, therefore it was not possible to measure the full apex motion accurately.

\subsection{Structural Flexure}

\begin{table}
\centering
\caption{Structural Flexure}
\begin{tabular}{|l|c|c|}
\hline
direction & VertexRSI & AEC \\
\hline
cross-elevation & 2.1 arcsec/(deg/s$^2$)& 1.6 arcsec/(deg/s$^2$) \\
elevation & 2.8 arcsec/(deg/s$^2$)& 2.1 arcsec/(deg/s$^2$) \\
\hline
\end{tabular}
\label{tab:flexure}
\end{table}

The pointing difference between encoders and accelerometers scales
well with the angular acceleration. The scaling factor is the antenna
stiffness for this type of load, and is summarised in Table \ref{tab:flexure}. Structure
flexure affects pointing accuracy during OTF scanning as a result of
large angular accelerations during scan reversal, which affect pointing by 0.8
arcsec in elevation, and 0.6 arcsec in cross-elevation for the
VertexRSI antenna, and 0.6 arcsec and 0.5 arcsec, respectively, for
the AEC antenna. Since the driving force for the deformation is an acceleration,
the numbers are also indicative for the stiffness of the structure for gravity
deformation, though the exact numbers there will be different and could not be
derived from the measurements performed during this investigation.

\subsection{Vibration Environment}

\subsubsection{Eigenfrequencies}

\begin{figure}
\resizebox{\hsize}{!}{
%\centering
\includegraphics[scale=0.4]{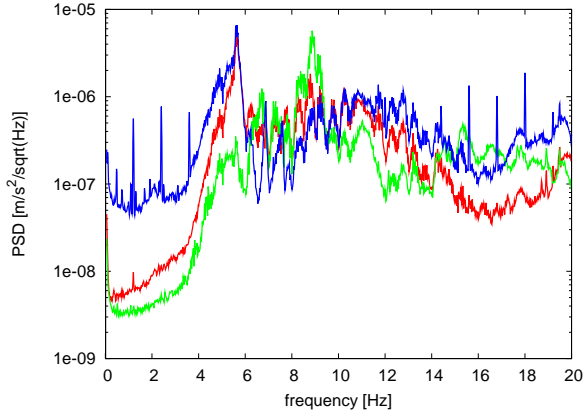}}
\caption{VertexRSI elevation (red), cross-elevation (green) and
  boresight (blue) motion power spectral density. The antenna was in
  shutdown mode at 45 degrees elevation during a typical day at the
  ATF. The curves for the elevation and cross-elevation motion are in
  units rad/s$^2$/$\sqrt(Hz)$. The spikes in the boresight motion curve
  are the result of vibrations of the receiver flange introduced by the
  cryogenic pump for the receivers.}
\label{fig:vertexeigen}
\end{figure}

\begin{figure}
\resizebox{\hsize}{!}{
%\centering
\includegraphics[scale=0.4]{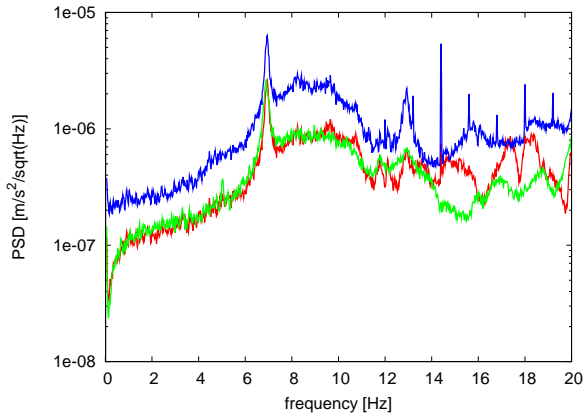}}
\caption{AEC elevation (red), cross-elevation (green) and
  boresight (blue) motion power spectral density. The antenna was in
  shutdown mode at 45 degrees elevation during a typical day at the
  ATF. The curves for the elevation and cross-elevation motion are in
  units rad/s$^2$/$\sqrt(Hz)$.}
\label{fig:aeceigen}
\end{figure}

Figure~\ref{fig:vertexeigen} illustrates some of the VertexRSI antenna
locked rotor eigenfrequencies, and Figure~\ref{fig:aeceigen}
illustrates some of the AEC antenna. The antennas were in shutdown
mode at approximately 45 degrees elevation. The figures show the
elevation, cross-elevation, and boresight motion of the BUS. The
lowest significant eigenfrequency is visible in elevation and
boresight at 5.57 Hz (VertexRSI antenna) and 6.8 Hz (AEC
antenna). There is a small peak visible at a frequency of 4.68 Hz for
the VertexRSI antenna, which is the print-through of the apex
structure rotation mode, along an axis through boresight.  

Some equipment installed on submillimeter radio telescopes, in
particular bolometers, may be sensitive to vibration. The
accelerometers provide a very accurate measure of the vibration
environment provided by the antenna. Two locations on the antenna were
investigated in some detail: the receiver flange, and the apex
structure. For both locations, 3-axis accelerometer measurements are
available for a variety of conditions.  

Under windy conditions, without sidereal tracking, the RMS
acceleration on the VertexRSI antenna receiver flange, combining all 3
axes, amounts to 0.48 mm/s$^2$. For the apex structure, the
corresponding number is 4.4 mm/s$^2$. For similar conditions, the AEC
antenna has vibration levels of 0.40 mm/s$^2$ at the receiver flange,
and 3.6 mm/s$^2$ at the apex structure. 

\begin{figure}
\resizebox{\hsize}{!}{
%\centering
\includegraphics[scale=0.4]{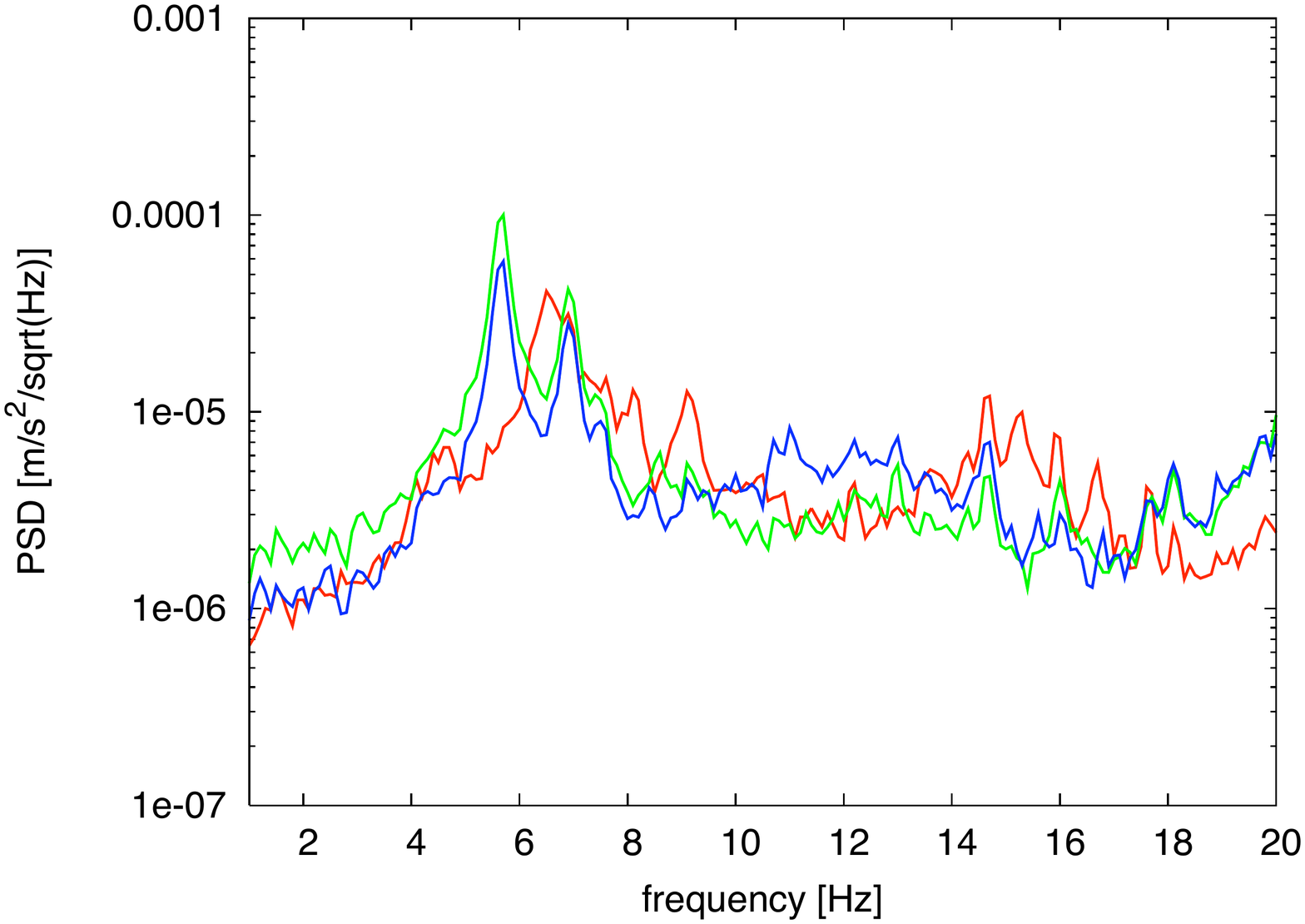}}
\caption{VertexRSI receiver flange acceleration PSD. Red, green and
  blue curves show the X, Y and Z components of the acceleration. Wind
  was approximately 9 m/s, and elevation was 45 degrees.}
\label{fig:vertexflange}
%\end{figure}
%\begin{figure}
\resizebox{\hsize}{!}{
%\centering
\includegraphics[scale=0.4]{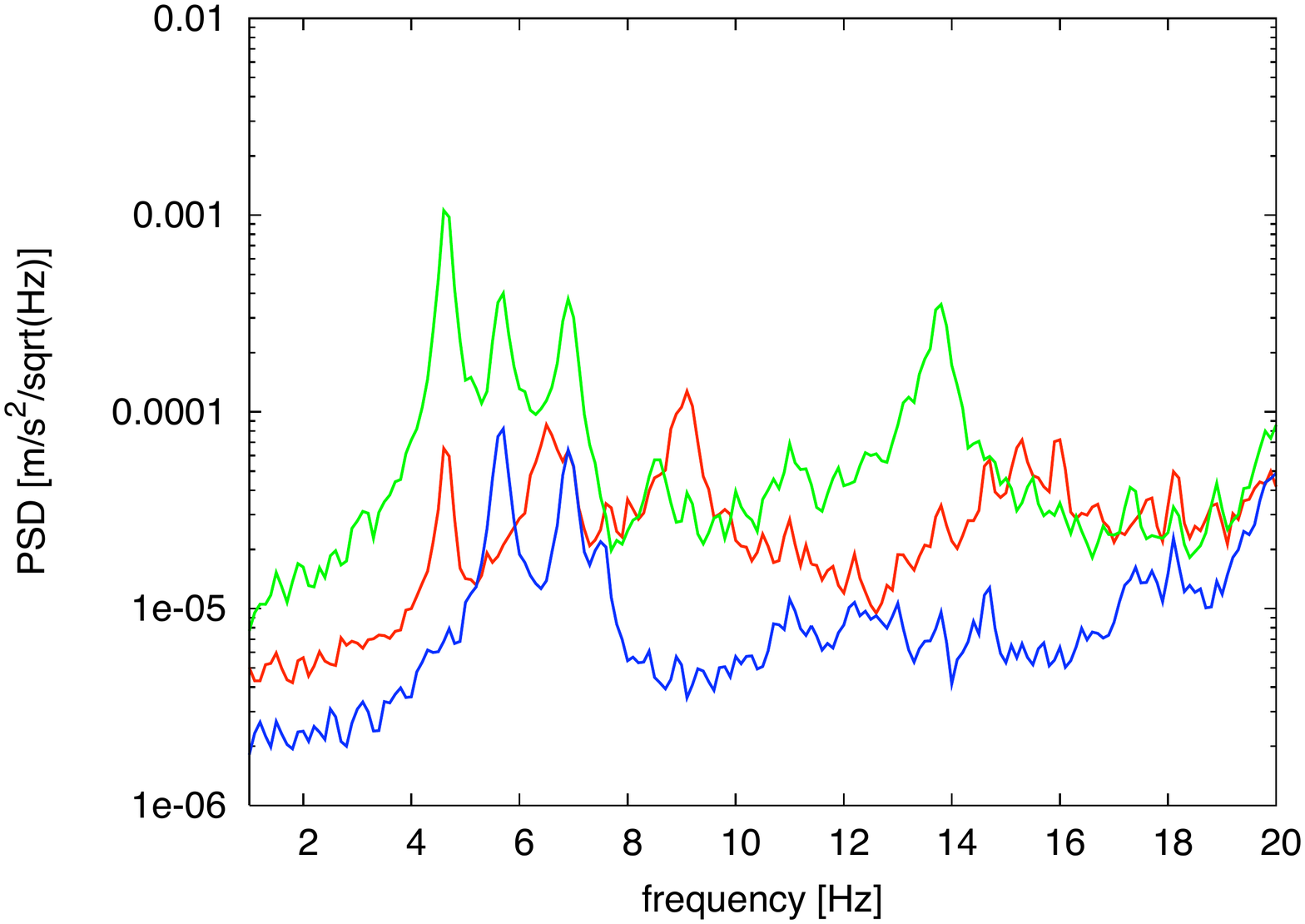}}
\caption{VertexRSI apex acceleration PSD for the same conditions as
  those shown in Figure~\ref{fig:vertexflange}.}
\label{fig:vertexapex}
\end{figure}

For the VertexRSI antenna, Figures~\ref{fig:vertexflange} and
\ref{fig:vertexapex} show the X (red, along the elevation axis), Y
(green, perpendicular to the elevation axis), and Z (blue, boresight)
PSDs of the acceleration, for the receiver flange and apex,
respectively. The apex structure rotates slightly around its own axis,
with a frequency of approximately 4.7 Hz. The amplitude of the
rotation is small, but affects the vibration environment depending on
the location of the accelerometer. For this specific measurement, the
accelerometers were placed near the edge of the apex cylinder, and
makes the Y measurement sensitive to rotation as well as displacement.  

\begin{figure}
\resizebox{\hsize}{!}{
%\centering
\includegraphics[scale=0.4]{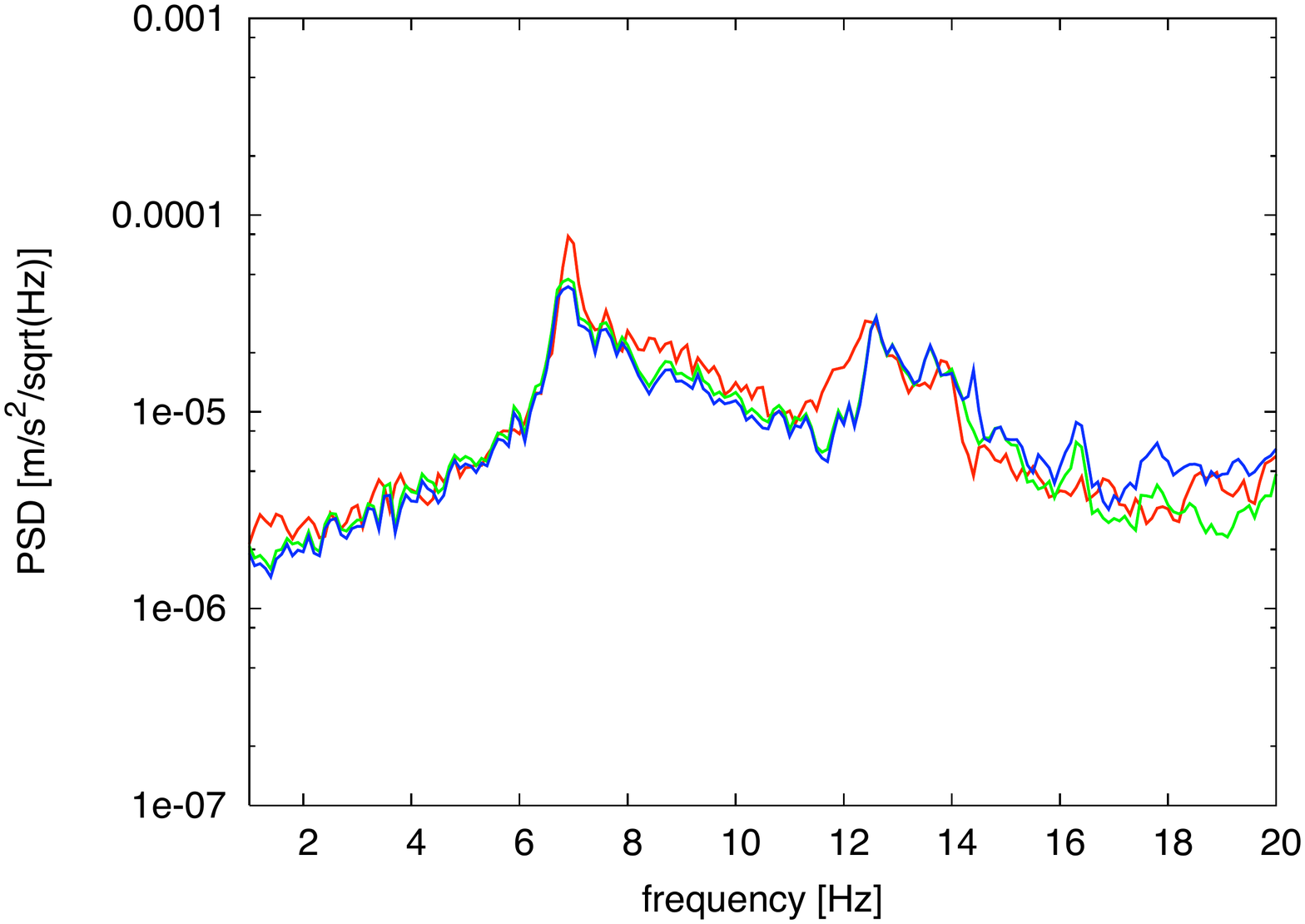}}
\caption{AEC receiver flange acceleration PSD. Red, green and
  blue curves show the X, Y and Z components of the acceleration. Wind
  was approximately 9 m/s, and elevation was 45 degrees.}
\label{fig:aecflange}
%\end{figure}
%\begin{figure}
\resizebox{\hsize}{!}{
%\centering
\includegraphics[scale=0.4]{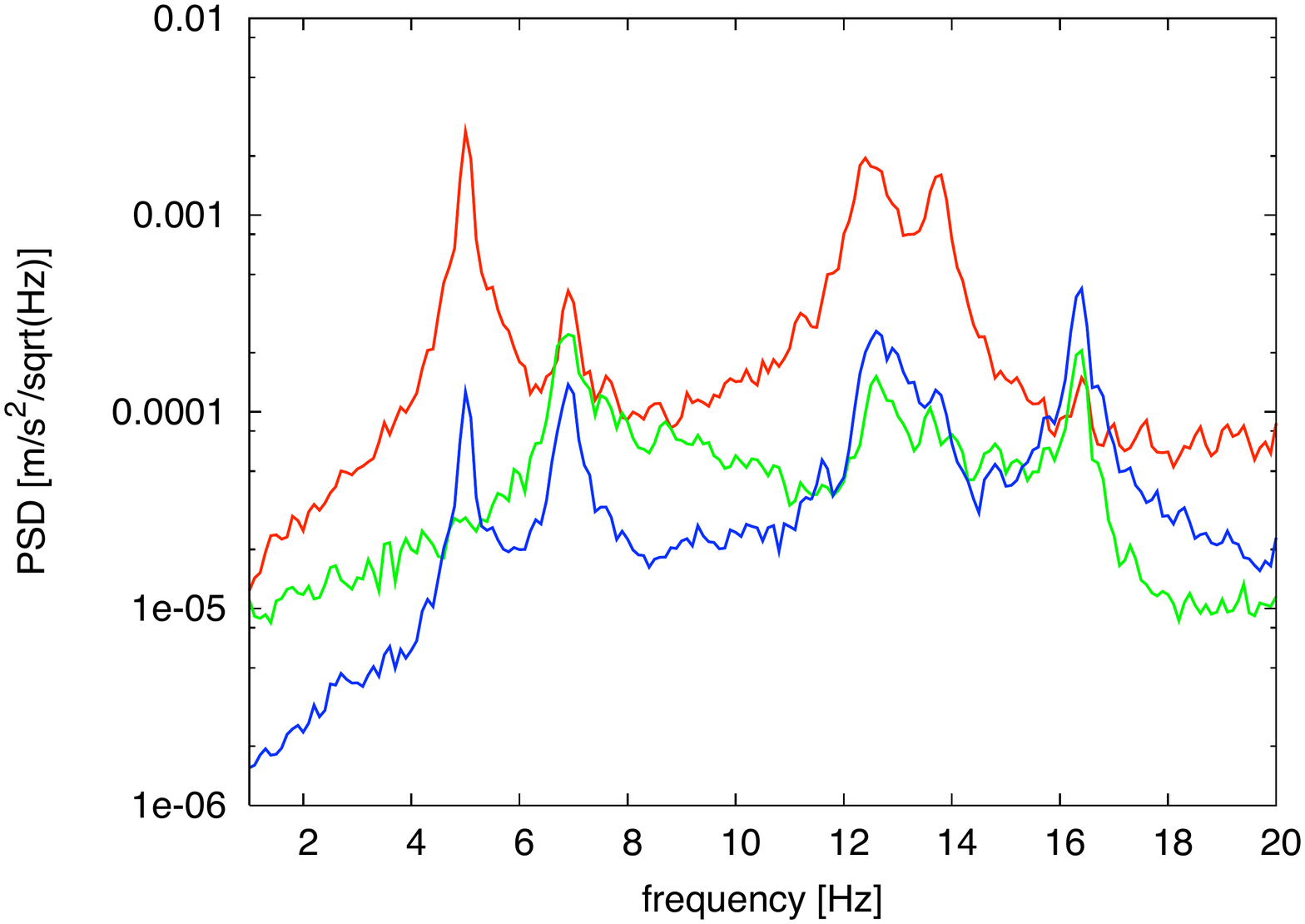}}
\caption{AEC apex acceleration PSD for the same conditions as those shown in Figure \ref{fig:aecflange}.}
\label{fig:aecapex}
\end{figure}

For the AEC antenna, Figures~\ref{fig:aecflange} and \ref{fig:aecapex}
show the X (red), Y (green), and Z (blue) PSDs of the acceleration,
for the receiver flange and apex, respectively. Also the AEC antenna
apex structure rotates somewhat around its own axis, with a frequency
of 5.0 Hz. For this specific measurement, the accelerometers were
placed on the outer part of the apex cylinder, and makes the X
measurement sensitive to rotation as well as displacement.  

\subsubsection{External Vibration Pick-Up}

During performance testing it became clear that the antennas are
sensitive to vibrations caused by motion of the other antennas, which
are transmitted through the ground. In order to test this sensitivity
in a more controlled way, the Mitsubishi antenna, located 35 m to the
west of the VertexRSI antenna, was instructed to make a 2 degree azimuth slew, while
accelerometers monitored the motion of the VertexRSI antenna. The move
caused peak-to-peak pointing errors of 0.30 arcsec in elevation, 0.40
arcsec in cross-elevation, and 2.9 $\mu$m in boresight motion. The
elevation for the VertexRSI antenna was 30 degrees. 

Vibration transfer from the VertexRSI antenna to the AEC antenna, also
placed 35 m apart, was
investigated by making the VertexRSI antenna perform fast switches
with 1 degree offset in both azimuth and elevation, and by making it
perform an interferometric OTF scan at 0.05 deg/s. The elevation of
the AEC antenna was 30 and 10 degrees respectively, during the
tests. The accelerometers mounted on the AEC antenna measured RMS
motion during the fast switching of 0.043 arcsec in elevation, 0.012
arcsec in cross elevation, and 0.7 $\mu$m RMS in boresight
motion. However, since the motion is peaked during the acceleration of
the VertexRSI antenna, and undetectable a few seconds after the move,
the peak-to-peak values of the pointing errors are of interest
too. Peak-to-peak motion was 0.29 arcsec in elevation, 0.13 arcsec in
cross-elevation, and 5.3 $\mu$m in boresight motion. For the
interferometric OTF scan, the numbers are 0.016 arcsec RMS or 0.23
arcsec peak-to-peak in elevation, 0.011 arcsec RMS or 0.31 arcsec
peak-to-peak in cross-elevation, and 0.29 $\mu$m RMS or 4.4 $\mu$m
peak-to-peak boresight motion.

These numbers are presented to give an impression of the impact of
the motion of nearby antennas. The tests were far too limited to
draw any further conclusions, and would depend critically on the soil
conditions at the ALMA and ATF sites.

\section{Discussion}

All wind-related antenna data obtained at the ATF have been analysed
in such a way as to represent antenna properties, independent of the
shape of the wind spectrum, or wind speed. With these antenna
properties, it is possible to predict wind-driven antenna performance
for any wind speed, wind spectrum, and air density, even for
conditions not encountered during antenna testing. One of the expected
and observed properties of the transfer functions as defined in Eqn. \ref{eq:psd}
allows a simple extrapolation of antenna properties measured between
0.1 Hz and a few Hz to values below 0.1 Hz, and in principle to 0
Hz. In the frequency domain this is a simple exercise, but in time
domain it corresponds to extrapolation of antenna properties measured
at timescales of seconds to timescales of tens of minutes or
longer. Since the wind contains the majority of its power at low
frequencies, this extrapolation is extremely useful for the prediction
of the overall wind performance of the antenna. 

Besides extrapolation to lower frequencies, the transfer functions can
be used to scale antenna performance measured under uncontrolled wind
conditions to any known and well-defined wind spectrum. The antennas
have been designed to meet the specifications for a given reference
spectrum, which was given in the SoW. In order to test compliance
with the specifications, the measured antenna performance must be scaled to this
SoW reference spectrum, a straightforward task, which does not leave
much room for speculation but provides hard numbers instead. 

One must realise, however, that the numbers for antenna performance
obtained using the wind transfer functions are valid only under the
assumptions made here, i.e. that wind effects dominate the
performance. For timescales of seconds to tens of seconds, this may
very well be the case, but when extrapolated to timescales of tens of
minutes to hours, other effects such as thermal effects may contribute
significantly to the total antenna performance at these timescales.

\subsection{Apex Rotation: Experiences with Accelerometer Placement
  and as Tool for Troubleshooting}

The results presented in this paper are for pointing stability as derived from
motion of the BUS only. The position of the subreflector determines
how the primary focus image is projected at the secondary focus, and
shake of the subreflector with respect to the BUS may introduce
additional pointing variations. 

During the measurements, it became clear that the apex structures of
both antennas rotate about the boresight axis. With only 3
accelerometers at the apex structure, it was no longer possible to
discern between rotations and translations of the subreflector.  At
the prime focus the plate scale is 34 arcsec/mm, which requires
stability of the apex structure to be on the order of a few tens of
$\mu$m. Using reconfigured accelerometers on one of the antennas, it was
possible to distinguish between rotation and translation in one
dimension. This revealed that for some resonance modes, the pointing
errors introduced by the BUS were compensated by those introduced due
to subreflector translation. Thus, the total pointing error projected
at the receiver flange may be larger or smaller than the pointing
stability derived from the BUS. 

Excessive rotation of the AEC antenna apex structure highlighted both
strengths and weaknesses of the accelerometer concept used in these
investigations. The main weakness was the inability of the
3-accelerometer set-up on the apex structure to discern between
translation and rotation of the structure. Since the apex structure
rotation was not foreseen at the time the accelerometer system was
designed, no provisions were made to distinguish between rotation and
translation. As it turned out, the rotation was so large that careful
on-axis placement of all three accelerometers would have been
necessary, not a trivial task given the dimensions of the
accelerometers. 

The strength of the accelerometer system was demonstrated when one of
the accelerometers at the apex was reconfigured to allow distinction
between rotation and translation (at the cost of the ability to
measure one displacement dimension). With the new configuration, it
could be determined that the detected accelerations were indeed caused
by rotation of the apex structure, that the rotation was off-axis, and
that the rotation axis shifted with elevation.  This off-axis rotation
of the apex structure translates into a pointing error of up to 1
arcsec peak-to-peak in cross-elevation, provided that the rotation
axis is parallel to the boresight axis (which could not be
confirmed).  In summary, this sequence of tests illustrated the
versatility of the accelerometer system as a diagnostic tool for
troubleshooting.

\subsection{Antenna Wake Turbulence: Spin-Off and Implications for
  Compact Array Configurations}

The SoW has a primary operating condition for antenna wake conditions,
with an average wind speed of 7 m/s instead of 9 m/s for non-wake
conditions, and a variable component of 4 times the SoW reference wind
spectrum. Analysis of wind data measured in the wake of the 3 antennas
placed at the ATF, and OPT pointing stability with the AEC antenna in
the wake of both the Mitsubishi and VertexRSI antennas, has revealed
detailed properties of the antenna wake turbulence.  

\begin{figure}
\resizebox{\hsize}{!}{
%\centering
\includegraphics[scale=0.4]{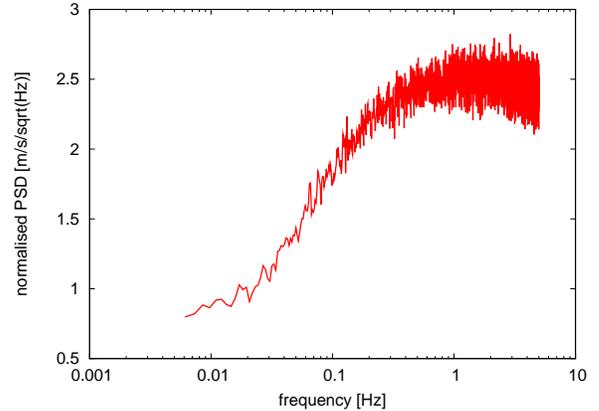}}
\caption{Antenna wake turbulence as measured with the sonic anemometer,
for wind passing the Mitsubishi prototype antenna. The PSD was scaled with
the wind speed, and divided by the average scaled PSD for wind coming
from unobstructed directions.}
\label{fig:wake_turbulence}
\end{figure}

\begin{figure}
\resizebox{\hsize}{!}{
%\centering
\includegraphics[scale=0.35, angle=90]{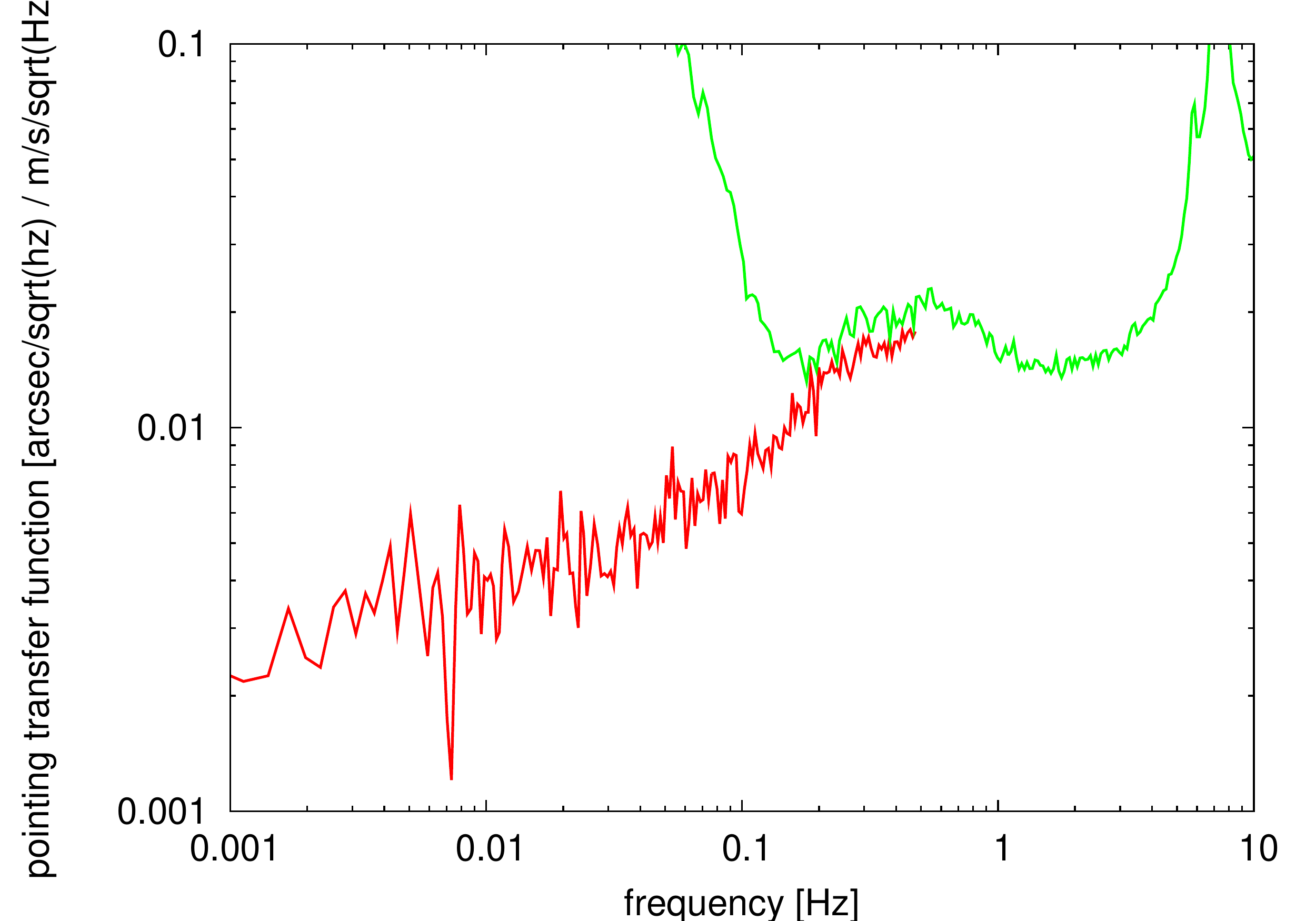}}
\caption{Antenna wake turbulence and AEC antenna resonance as measured with 
the OPT and accelerometers,
for wind passing the Mitsubishi and VertexRSI prototype antennae. The PSD
for elevation pointing was scaled with
the wind speed, and divided by the PSD of the wind
during the measurement.
The green curve is measured with the accelerometers, the red curve
with the optical pointing telescope. Wind was approximately 12 m/s average.}
\label{fig:OPT_wind_stiffness}
\end{figure}

Figure~\ref{fig:wake_turbulence} shows the effect antenna wake
turbulence has on the undisturbed wind, as measured with the sonic
anemometer, for wind passing the Mitsubishi prototype antenna. The
measured PSD downwind of the antenna was scaled with the average wind
speed, and divided by the average scaled PSD for wind coming from
unobstructed directions. The curve shows how the unobstructed wind is
affected by the antenna, at a distance of approximately 30 m. The
scaling factor for the high frequencies is approximately flat with
frequency, at a value of 2.5, but the low frequency part of the
spectrum is not affected by the same amount. Turn-over occurs in the
range around 0.1 Hz. This means that the turbulence introduced in the
antenna wake has frequencies of 0.1 Hz and higher, and will at these
frequencies shake any antenna placed in this wake a factor 2.5 more
than in the absence of the extra turbulence. 

Figure~\ref{fig:OPT_wind_stiffness} shows the antenna wake turbulence
as measured on the AEC antenna with the accelerometers and the OPT
simultaneously. The wind was predominantly from the west, passing over
the Mitsubishi and VertexRSI antennas. The high frequency part of the
plot is dominated by antenna resonances and vortex shedding off of the
other antennas, measured with the accelerometers. Below the lowest
eigenfrequency around 7 Hz, the curve is expected to flatten, which is
also seen down to frequencies of about 0.2 Hz. Below this frequency,
the accelerometers are affected by noise. The simultaneous
measurements with the OPT, tracking on Polaris, show an overlap up to
0.7 Hz, and are not affected by low frequency noise, and thus valid
down to the lowest frequencies.  

Note the good overlap of the curves in the frequency range between 0.2
and 0.7 Hz, as expected under the assumption that both the OPT and
accelerometers see the same BUS motion. The effects of combined
Mitsubishi and VertexRSI turbulence are also here clearly visbible,
with a turn-over frequency around 0.1 Hz. The magnitude of the high
frequency turbulence gain is approximately a factor 4 to 5.  

Thus the SoW requirement for 4 times the wind spectrum (which
translates to 2 times the wind spectrum in the units used in
the figures here, m/s/$\sqrt{Hz}$) in the wake of an 
antenna is valid, but only for frequencies above 0.1 Hz.  It is also
clear from the ATF antenna wake turbulence, that the level of the
turbulence depends on the distance to the antenna, and that the
effects of multiple antennae appear to be multiplicative.

Since the low frequency (0.001 Hz) antenna wind performance is
dominated by the low frequency stiffness of the antenna for wind
excitation, the resonances, which are all well above 1 Hz, and the
wake turbulence generated by nearby antennas, at frequencies above
approximately 0.1 Hz, do not significantly affect wind-related antenna
performance for timescales of 15 minutes (0.001 Hz). The cumulative
effect of several antennas upwind may, however, not be negligible as
seen with the combined Mitsubishi and VertexRSI wake turbulence. Nor
will the wake turbulence be negligible any longer when a properly
functioning metrology system suppresses the low frequency (below 0.1
Hz) wind buffeting of the antennas, or when antenna performance
(beyond the requirements) at frequencies above 0.1 Hz is an issue.

\section{Conclusions}

\subsection{Accelerometers}

Seismic accelerometers mounted on the back-up structures of large
reflector antennas are capable of characterising all relevant BUS
rigid body motions, and a few of its low-order deformation
modes. Independently of any external sources to the antenna, and even
for antennas without receivers, it is possible to derive the dynamical
behaviour of the performance parameters of an antenna, such as
pointing accuracy, primary reflector surface stability, and path
length stability. The accuracy at which performance parameters can be
measured is a function of frequency, and typically at the sub-$\mu$m
and sub-arcsecond level for frequencies above 0.1 Hz. 

In combination with wind measurements, antenna performance for windy
conditions can safely be extrapolated to frequencies well below the
limit of 0.1 Hz imposed by noise in the accelerometer data. The
validity of this extrapolation has been demonstrated with the use of
an optical pointing telescope, not limited by the low-frequency
noise. 

In addition to performance testing, the robust accelerometer system
has proven to be well suited for troubleshooting of unexpected antenna
behaviour, such as large and off-axis apex structure rotation for the
AEC antenna, and servo-tuning issues for the VertexRSI antenna. 

\subsection{Wind-Driven Performance}

Performance of the antennas for windy conditions was a major design
driver. In spite of the very different wind conditions at the antenna
test site and the ALMA site, it was possible to accurately predict
performance for the ALMA site based on measurements performed at the
ATF. The key to this extrapolation is careful characterisation of the
wind at the sites, in particular at the ATF site. Wind-driven
performance as measured on the antennas was combined with the wind
characteristics during the time of measurement, which allowed
calculation of wind-independent antenna properties, allowing
calculation of antenna performance for any wind condition.

As a spin-off of the investigations, antenna wake turbulence at a
typical distance of 30 - 50 m was determined; which is useful
information for calculation of antenna performance in the compact
configuration.

\subsection{Antenna Performance}

Antenna performance as measured with the accelerometers complements
other performance measurements, such as  optical and radio pointing,
and holography. The full performance of the antennas could not be
determined with any of the individual methods, but a combination of
them gives confidence in the completeness of the test results. 

During design, antenna performance was calculated from the sum of
individual contributions to the total error budget. The performance
numbers presented in this paper are best interpreted in the context of
the corresponding error budget contributions used for antenna
design. Table~\ref{tab:pointing} gives the measured performance for
each antenna, and the calculated error budget entry as taken from the
design documentation where available
\cite{Mangum2006}. %% RS 27Nov2006

\begin{table*}
\centering
\caption{Wind Pointing}
\begin{tabular}{lll}
& VertexRSI & AEC \\
\\
Pointing accuracy (wind only) & 0.81 arcsec& 0.45 arcsec\\
Pointing accuracy (wind only) error budget & 0.035 arcsec& 0.35 arcsec\\
Pointing accuracy (wind + tracking) & 0.94 arcsec& 0.50 arcsec\\
Pointing accuracy offset pointing requirement& 0.6 arcsec& 0.6 arcsec\\
\\
Primary reflector surface stability, wind effects (astigmatism + %defocus
AAM, at edge of BUS) & 5.3 + 2.2 $\mu$m& 6 + 5 $\mu$m\\
Primary reflector surface stability, wind effects error budget& 8.4  $\mu$m& 2.1  $\mu$m\\
Primary reflector surface stability, overall requirement& 25  $\mu$m& 25  $\mu$m\\
\\
Path length stability, wind effects & 6 $\mu$m& 6 $\mu$m\\
Path length stability, wind effects error budget& 7.6 $\mu$m& 3.5 $\mu$m\\
Path length stability, requirement total non-repeatable residual delay & 15 $\mu$m& 15 $\mu$m\\

\\
Structure flexure cross-elevation & 2.1 arcsec/(deg/s$^2$)& 1.6 arcsec/(deg/s$^2$)\\
Structure flexure elevation & 2.8 arcsec/(deg/s$^2$)& 2.1 arcsec/(deg/s$^2$)\\
\\
Lowest eigenfrequencies & 5.57 Hz & 6.8 Hz \\

\end{tabular}
\label{tab:pointing}
\end{table*}

% use section* for acknowledgement
\section*{Acknowledgment}
% optional entry into table of contents (if used)
%\addcontentsline{toc}{section}{Acknowledgment}
The authors would like to thank Nobuharu Ukita (National Astronomical
Observatory of Japan) and David R. Smith (Merlab) for valuable
discussions on the design of the accelerometer system and analysis of
the measurements, Angel Otarola and Juan Pablo Perez Beaupuits (ESO)
for providing wind data for the ALMA site, and Fritz Stauffer and
Nicholas Emerson (NRAO) for valuable support at the ATF. 

% trigger a \newpage just before the given reference
% number - used to balance the columns on the last page
% adjust value as needed - may need to be readjusted if
% the document is modified later
%\IEEEtriggeratref{8}
% The "triggered" command can be changed if desired:
%\IEEEtriggercmd{\enlargethispage{-5in}}

% references section
% NOTE: BibTeX documentation can be easily obtained at:
% http://www.ctan.org/tex-archive/biblio/bibtex/contrib/doc/

% can use a bibliography generated by BibTeX as a .bbl file
% standard IEEE bibliography style from:
% http://www.ctan.org/tex-archive/macros/latex/contrib/supported/IEEEtran/bibtex
%\bibliographystyle{IEEEtran.bst}
% argument is your BibTeX string definitions and bibliography database(s)
%\bibliography{IEEEabrv,../bib/paper}
\bibliographystyle{IEEEtran}
\bibliography{IEEEabrv,mybibfile}

\begin{thebibliography}{1}

\bibitem[1]{Kaercher2000} H.~J.~K\"archer and J.~W.~M.~Baars, ``The
  design of the Large Millimeter Telescope / Gran Telescopio
  Milimetrico (LMT/GTM)'', \textit{Proceedings SPIE}, \textbf{4015},
  2000, pp. 155-168. 

\bibitem[2]{Baars1973} J.~W.~M.~Baars, ``The Measurement of large
  Antennas with cosmic Radio Sources'', \textit{IEEE Trans.~Antennas
    Propagation}, \textbf{21}, 1973, pp. 461-474. 

\bibitem[3]{Mangum2006} J.~G.~Mangum, J.~W.~M.~Baars, A.~Greve,
  R.~Lucas, R.~Snel, P.~T.~Wallace and M.~Holdaway, ``Evaluation of
  the ALMA Prototype Antennas'', \textit{Publ.~Astron.~Soc.~of the
    Pacific}, \textbf{118}, 2006, pp. 1257-1301. 

\bibitem[4]{Smith2004} D.~R.~Smith, P.~Avitabile, G.~Gwaltney, M.~Cho
  and M.~Sheehan, ``Wind-Induced Structural Response of a Large
  Telescope'', \textit{Proceedings SPIE}, \textbf{5495}, 2004,
  p. 258

\bibitem[5]{Ukita2002} N.~Ukita and M.~Ikeda, ``Antenna Vibration
  Measurements with Accelerometers'', 2002, URSI General Assembly
  (Maastricht), 2002, p. 1958.

\bibitem[6]{Ukita2004} N.~Ukita, M.~Saito, H.~Ezawa, B.~Ikenoue,
  H.~Ishizaki, H.~Iwashita, N.~Yamaguchi, T.~Hayakawa, ``Design and
  performance of the ALMA-J prototype antenna'', \textit{Proceedings
    SPIE}, \textbf{5489}, 2004, pp. 1085-1093. 

\bibitem[7]{Zernike1934} F.~Zernike, ``Diffraction Theory of the
  Knife-Edge Test and its Improved Form, The Phase-Contrast Method'',
  \textit{Monthly Not.~Roy.~Astron.~Soc.}, \textbf{94}, 1934, p. 377 

\bibitem[8]{Greve2006} A.~Greve and J.~G.~Mangum, ``Mechanical
  Measurements of the ALMA Prototype Antennas'', \textit{IEEE Antennas
    and Propagation Magazine}, submitted 2006. 

\bibitem[9]{Davenport1961} A.~G.~Davenport, ``The Application of
  Statistical Concepts to the Wind Loading of Structures'',
  \textit{Proc.~of the Institute of Civil Engineers}, paper no. 6480,
  1961, pp. 449-472.


\end{thebibliography}
%
% <OR> manually copy in the resultant .bbl file
% set second argument of \begin to the number of references
% (used to reserve space for the reference number labels box)

\appendices

\section{Mathematical Treatment of Accelerometer Signals}
%================
\label{education}

The mathematical treatment of the accelerometer data as covered in sections 
\ref{setup} and \ref{methods} is elaborated in this appendix.
Accelerometer measurements yield the acceleration as a function of time, as 
measured at the location of the accelerometer with the direction along the
sensitive axis of the accelerometer.

In order to obtain a displacement signal from the measured accelerations, the
signal needs to be integrated twice, since the acceleration $A(t)$ is by 
definition the second time $t$ derivative of the displacement or position 
$X(t)$:

\begin{equation}
A(t) = \frac{d^2X(t)}{dt^2}
\end{equation}

Thus, for a given acceleration time series, the displacement becomes:

\begin{equation}
X(t) = \int\int A(t) d^2t
\end{equation}

where two integration constants need to be used. The integration constants are 
chosen in a way to minimise the mean and the slope of $X(t)$.

Many of the results in this paper are presented in the form of 
power spectra. The power spectrum of a time series $A(t)$ with $n$ samples
is defined by Equation~\ref{eq:aoft}, repeated here:

\begin{equation}
\hat{A}(\nu) = \frac{|FFT(A(t))|^2}{n^2}
\end{equation}

where $FFT$ is the fast Fourier transform. The normalisation with $n^2$ gives
the power spectrum the property that the sum of all terms of $\hat{A}(\nu)$
equals the variance of the time series $A(t)$.

The power spectrum of a doubly integrated acceleration time series becomes thus:

\begin{equation}
\hat{X}(\nu) =  \frac{|FFT( \int\int A(t)  d^2t )|^2}{n^2}
\end{equation}

Consider now the special case where $X(t) = \sin{2 \pi\nu_0 t}$, a harmonic motion
with frequency $\nu_0$. The second time derivative of $X(t)$ then becomes:

\begin{eqnarray}
A(t) &=& \frac{d^2 \sin{2 \pi\nu_0 t}}{dt^2} \nonumber \\
     &=& -(2 \pi\nu_0)^2 \sin{2 \pi\nu_0 t} \nonumber \\
     &=& -(2 \pi\nu_0)^2 X(t)
\end{eqnarray}

which is equal to the time series $X(t)$ times a factor $-(2 \pi\nu_0)^2$.

When the power spectrum is taken on either side of the equation, it results in 
the following expression:

\begin{eqnarray}
\hat{A}(\nu) &=& \frac{|FFT(-(2 \pi\nu_0)^2 X(t))|^2}{k^2} \nonumber \\
             &=& (2 \pi\nu_0)^4 \hat{X}(\nu)
\end{eqnarray}

where $|-(2 \pi\nu_0)^2|^2$ can be taken outside the Fourier transform since it 
is not a function of $t$. The remaining term is exactly the power spectrum of
$X(t)$. Division by $(2 \pi\nu_0)^4$ on either side of the equation gives:

\begin{equation}
\hat{X}(\nu) = \frac{\hat{A}(\nu)}{(2 \pi\nu_0)^4}
\end{equation}

Since any function can be described as a sum of sine functions with 
different amplitude, phase, and frequency, the frequency $\nu_0$ in $(2
\pi\nu_0)^4$ can be replaced with $\nu$, the coordinate in the frequency
domain of the power spectrum.

All power spectral densities (PSDs) presented in this paper are the square
root of the power spectrum defined above, thus the PSD of a timeseries of
accelerations, including double integration to obtain displacement, is
given by:

\begin{equation}
PSD(X(t)) = \sqrt{\hat{X}(\nu)} = \frac{\sqrt{\hat{A}(\nu)}}{(2 \pi\nu)^2} = \frac{PSD(A(t))}{(2 \pi\nu)^2}
\end{equation}

% biography section
% 
% If you have an EPS/PDF photo (graphicx package needed) extra braces are
% needed around the contents of the optional argument to biography to prevent
% the LaTeX parser from getting confused when it sees the complicated
% \includegraphics command within an optional argument. (You could create
% your own custom macro containing the \includegraphics command to make things
% simpler here.)
%\begin{biography}[{\includegraphics[width=1in,height=1.25in,clip,keepaspectratio]{mshell}}]{Michael Shell}
% where an .eps filename suffix will be assumed under latex, and a .pdf suffix
% will be assumed for pdflatex; or if you just want to reserve a space for
% a photo:

\begin{biography}[{\includegraphics[width=1in,height=1.25in,clip,keepaspectratio]{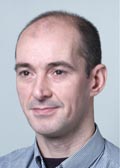}}]{Ralph C.~Snel} received his M.Sc. degrees in
  astrophysics and experimental physics at the University of Utrecht
  in the Netherlands in 1991 and 1992, and his Ph.D. degree in
  astrophysics at Lund University in Sweden in 1998. 

From observing steller populations with
the Hubble Space Telescope he turned his view downward and became
involved as calibration scientist for the Earth atmosphere observing
imaging spectrogrograph SCIAMACHY on the ESA Envisat satellite, at
SRON Space Research in the Netherlands. From 2001 to 2003 he worked at
Lund Observatory on development of an optical telescope in the 50 m
class, and testing of the ALMA prototype antennas, which he continued
under ESO and NRAO contracts in 2004. Recently, he joined SRON again
for more Earth observation. In his spare time, he is often found at
the anvil blacksmithing reconstructions of mediaeval arms and armour. 
\end{biography}

% if you will not have a photo at all:
\begin{biography}[{\includegraphics[width=1in,height=1.25in,clip,keepaspectratio]{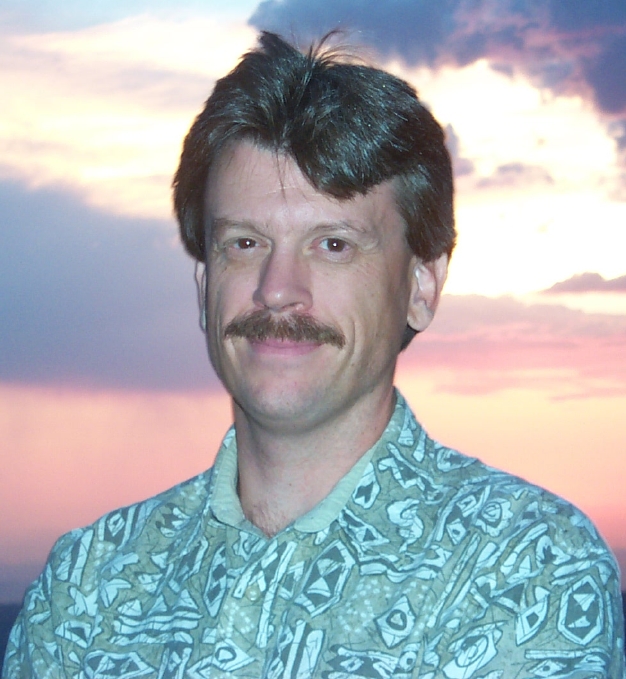}}]{Jeffrey G.~Mangum}
received the Ph.D. degree in astronomy from the University of Virginia
in 1990.  

Following a two-year residency as a postdoctoral researcher in the
astronomy department at the University of Texas he joined the staff of
the Submillimeter Telescope Observatory (SMTO) at the University of
Arizona.  In 1995 he joined the scientific staff at the National Radio
Astronomy Observatory (NRAO) in Tucson, Arizona, and subsequently moved to
the NRAO headquarters in Charlottesville, Virginia.  His research
interests include the astrophysics of star formation, the solar
system, and external galaxies, the performance characterization of
reflector antennas, and calibration of millimeter-wavelength
astronomical measurements.
\end{biography}

% insert where needed to balance the two columns on the last page
%\newpage

\begin{biography}[{\includegraphics[width=1in,height=1.25in,clip,keepaspectratio]{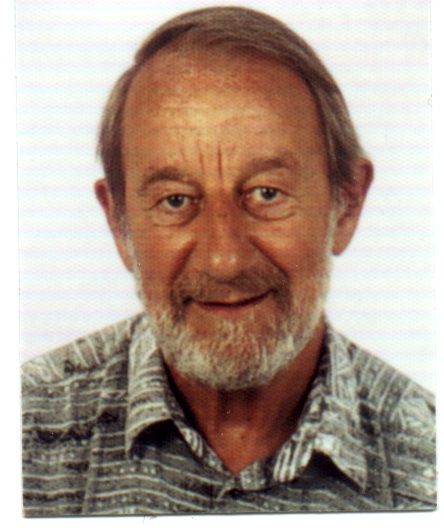}}]{Jacob W.~M.~Baars} received the M.Sc. and
  D.Sc. degrees in applied physics from the Technical University of
  Delft, the Netherlands in 1963 and 1970, respectively. 

After working at the National Radio
Astronomy Observatory in Green Bank, WV, he joined the Netherlands
Foundation for Radio Astronomy in 1969. There he participated in the
construction and was later head of the Westerbork Synthesis Radio
Telescope. In 1975 he joined the Max-Planck-Institut f\"ur
Radioastronomie in Bonn as head of the Division for Millimeter
Technology. He was project Manager of the 30-m Millimeter Radio
Telescope of IRAM in Spain and the Heinrich Hertz Telescope in
Arizona. From 1997-99 he worked on the UMass-Mexico Large Millimeter
Telescope. In 1999 he joined the European Southern Observatory, where
he was involved in several aspects of the ALMA Project, lastly the
evaluation of the prototype antennas. Since his retirement he consults
in the area of large antennas and radio telescopes. His research
interests are in the area of antenna theory and practice, in
particular the performance calibration of large antennas with radio
sources, and of atmospheric influences on observations at very high
frequencies. 
\end{biography}

% You can push biographies down or up by placing
% a \vfill before or after them. The appropriate
% use of \vfill depends on what kind of text is
% on the last page and whether or not the columns
% are being equalized.

%\vfill

% Can be used to pull up biographies so that the bottom of the last one
% is flush with the other column.
%\enlargethispage{-5in}

% that's all folks

\end{document}